\documentclass[letterpaper,11pt]{article}
\pdfoutput=1 

\usepackage{jheppub} 

\usepackage{tikz}
\usepackage{dsfont}
\usepackage{amsmath,empheq}
\def\ba{\begin{align}}\def\ea{\end{align}}

\definecolor{darkgreen}{rgb}{0,0.4,0}
\def\beq{\begin{eqnarray}}\def\eeq{\end{eqnarray}}
\def\be{\begin{equation}}\def\ee{\end{equation}}
\def\ben{\begin{equation}}
\def\een{\end{equation}}
\def\bea{\begin{eqnarray}}
\def\eea{\end{eqnarray}}

\def\vev#1{\langle{#1}\rangle}

\newcommand{\nn}{\nonumber}
\newcommand{\mcA}{\mathcal{A}}
\newcommand{\tmcA}{\tilde{\mathcal{A}}}
\newcommand{\mcP}{\mathcal{P}}

\def\t6 {T_\mt{D6}}

\newcommand{\mt}[1]{\textrm{\tiny #1}}

\def\cale         {{\cal E}}

\def\ee           {{\rm e}}

\def\sqr#1#2{{\vcenter{\vbox{\hrule height.#2pt
 \hbox{\vrule width.#2pt height#1pt \kern#1pt
 \vrule width.#2pt}\hrule height.#2pt}}}}

\def\ee{\cale}

\def\aa1{\phi}
\def\cc1{\psi}

\def\vev#1{\langle #1 \rangle}

\def\ket#1{|#1\rangle}

\def\vev#1{\langle{#1}\rangle}

\def\ndt{\noindent}
\def\nd{{ \vphantom{\dagger}}}

\def\eg{{\it e.g.}}

\title{Chaotic and Thermal Aspects in the Highly Excited String S-Matrix}

\author{Diptarka Das,}
\author{Santanu Mandal,}
\author{Anurag Sarkar}

\affiliation{ Department of Physics, Indian Institute of Technology, Kanpur, UP 208016, INDIA.}
\emailAdd{didas@iitk.ac.in}
\emailAdd{sarkara@iitk.ac.in}
\emailAdd{santanum20@iitk.ac.in}

\abstract{We compute tree level scattering amplitudes involving more than one highly excited states and tachyons in bosonic string theory. We use these amplitudes to understand the chaotic and thermal aspects of the excited string states lending support to the Susskind-Horowitz-Polchinski correspondence principle. The unaveraged amplitudes exhibit chaos in the resonance distribution as a function of the kinematic parameters, which can be described by random matrix theory. Upon coarse-graining, these amplitudes are shown to exponentiate, and capture various thermal features, including features of a stringy version of the eigenstate thermalization hypothesis as well as notions of typicality. Further, we compute the effective string form factor corresponding to the highly excited states, and argue for the random walk behaviour of the long strings.}

\begin{document}

    \begin{flushleft}
\end{flushleft}  

\maketitle
\flushbottom
\vspace{10pt}

%
%
%
%
%
%
%
%
%
%
%
%
%
%
%
%



\section{Context}

{\bf Black holes are chaotic and thermal quantum objects.} Both these notions have by now been well established classically. For instance, it has been known since long that classical orbits around Schwarzchild black holes exhibit chaos \cite{Bombelli:1991eg}. The quantum origins of these phenomena is an area of active research, with connections being recently made between black holes and quantum chaos in \cite{Shenker:2013pqa, Shenker:2013yza, Leichenauer:2014nxa, Shenker:2014cwa}. The horizon plays an important role in these computations which gives rise to exponential red-shift. The quantum effects of horizon physics can be traced back to Hawking's calculation of black hole radiation \cite{Hawking:1975vcx} which points towards a {\em thermal} interpretation. This has since then created the information loss enigma. In a full quantum gravity theory, this problem can be cast as a feature of the theory, known as the {\it central dogma} \cite{Almheiri:2020cfm} : which states that the black hole as seen by an exterior observer is described by a unitary theory with $A/4G$ quantum degrees of freedom. A strong evidence for the dogma comes from reproduction of the Bekenstein-Hawking  black hole entropy formula for special extremal black holes in supersymmetric string theories \cite{Strominger:1996sh, Callan:1996dv}. This counting is possible in the string theory's weak coupling limit due to enough supersymmetry. A primer to these computations in the context of black holes in string theory appear through the {\it correspondence principle}. 

\noindent
{\bf The Susskind-Horowitz-Polchinski correspondence principle} \cite{Susskind:1993ws, Horowitz:1996nw} conjectures that the Schwarzchild black hole is adiabatically connected to a single free string state. For {\it freeness} one need a highly excited state of the string, $\ket{HES}$,  i.e. at a large level $N \gg 1$ so that, $M_H = \sqrt{ N/\alpha'}$ is large. Since the phases are identified this is also the mass of the Schwarzchild black hole, $M$. At the correspondence point, the string length $\ell_s$ equals the Schwarzchild radius of the black hole $2 \,G M = 2 \,g^2 \ell_s^2 M_H$, where $G$ is the Newton's constant expressed in terms of $g$, the string coupling. Equality of the length scales chooses a special value of the string coupling, $g = N^{-1/4}$. Hence we see that large $N$ implies, free strings. Upto the factor of $4$, the correspondence point also reproduces the Bekenstein-Hawking formula. 

\noindent
Amati and Russo \cite{Amati:1999fv} (see also \cite{Cornalba:2006hc, Iengo:2003ct, Chen:2005ra, Chialva:2004xm, Iengo:2006if, Matsuo:2009sx, Manes:2001cs, Kuroki:2007aj}) discovered blackbody spectrum by considering {\it coarse-grained} decay amplitude of $\ket{HES}$ going to another $\ket{HES}$ and a tachyon/photon. The coarse-graining process involved averaging over initial HES states and summing over final ones, keeping level (mass) fixed. The temperature of the radiation coincided with the Hagedorn temperature $1/\sqrt{\alpha'}$ as expected at the correspondence point. This calculation set-up has in-built coarse-graining and is insensitive to the fine-grained dynamics of individual microstates. It is however in the microstate dynamics, where the quantum origins of chaos lie. 

\ndt
{\bf Probing quantum chaos in the  black hole S-matrix. } A measure of classical chaos is sensitivity of dynamics to the initial conditions. In quantum case the main observables are correlation functions and S-matrices. Through the former, quantum chaos is captured via the exponential time dependence in out-of-time-ordered-correlators \cite{larkin1968quasi}. In the S-matrix too, its sensitivity to changes of the scattering microstate gives a notion of chaos \cite{ott1993chaotic, Rosenhaus}. In chaotic scattering, the resonance peak positions in the S-matrix amplitude themselves start showing random matrix statistics, as one changes the initial microstate slightly. In particular the positions of the S-matrix peaks (as a function say of the scattering angle) can be interpreted as eigenvalues of a random matrix. Typically, this is the case when the scattering involves highly excited states or classically chaotic potentials. One also expects chaos in the black hole S-matrix \cite{Polchinski}. Therefore it is natural to look for indications of chaotic scattering if one can compute the fine-grained string S-matrix involving $\ket{HES}$.  Such a computation is facilitated via the Del Giudice, Di Vecchia, Fubini (DDF) states \cite{DelGiudice:1971yjh}. 

\ndt{\bf Highly excited DDF states} are created by hitting a tachyon vertex operator with series of photon vertex operators and successively picking out the intermediate states appearing in the OPEs through contour integrals. The construction algorithmizes the Virasoro constraint, making the states physical. In addition to the spacetime momentum $p$, these states are also labelled by a polarization vector $\zeta$ which is inherited from the series of photons involved in its construction. Scattering amplitudes involving these states were computed in \cite{GR, BF1, Firrotta:2024qel}. Explicitly the amplitude involving a single $HES(p,\zeta)$ and two tachyons $T(p)$ were evaluated, the tachyons were generalized to photons in \cite{FR}. A concrete proposal to extract chaotic features from these amplitudes was proposed in \cite{BF-PRL}. The  analysis was carried out in the amplitude involving a $HES$ and few tachyon scattering amplitudes in \cite{BF2}. \footnote{In \cite{Hashi} another proposal was analyzed in String-scattering which involved looking for fractals in $2\leftrightarrow 2$ scattering amplitudes. } 

\ndt In this paper we evaluate scattering amplitudes involving more than one $\ket{HES}$ states of the DDF type. In particular we consider the following two S-matrices : 
\begin{itemize}
 \item  A $\ket{HES}$ decaying into a tachyon and another $\ket{HES}$
\begin{align}
HES_1(p_1,\zeta_1) \rightarrow HES_2(p_2, \zeta_2) + T_3(p_3)
\label{eq1}
\end{align}
\item  A $2\leftrightarrow 2$ scattering of a $\ket{HES}$ and a tachyon going into another $\ket{HES}$ and a tachyon. 
\begin{align}
HES_1(p_1,\zeta_1)+T_2(p_2)  \to HES_3(p_3,\zeta_3)+T_4(p_4)
\label{eq2}
\end{align}
\end{itemize}
The expressions for the $HHT$ and the $HHTT$ amplitudes are derived in Eq.\eqref{HHT amp} and in Eq.\eqref{hhtt-amp} respectively. In the rest of the paper we use these amplitudes as probes to understand thermalization and chaos. Our discussions can be arranged along various axes
\begin{enumerate}
\item {\it What are the signatures of chaos in $HES$ scattering? }  The S-matrix as a function of the scattering angle exhibits peaks. The position of the peaks are extremely sensitive to the microstate constituency of the both the initial as well as the final $\ket{HES}$ states. We followed the numerical analysis of \cite{BF2} that was carried out for the $HTT$ amplitude, to uncover the chaotic distribution of peaks in the S-matrix as a function of the scattering angle. 
\begin{enumerate}
\item {\em Three point $HHT$ amplitude chaos: } In \S\ref{subsec:HHT-chaos-eth} we evaluated $HHT$ numerically and compared the peak statistics with random matrix theory (RMT). Level-repulsion is clearly evident, and we compared the level statistics distribution of peaks with that which follows from the Gaussian Orthogonal Ensemble (GOE) universality class of RMT. We do this by comparing the two probability distributions using a one parameter fit as well as the Kolmogorov-Smirnov test. Our results indicate that as the level of the $\ket{HES}$ states increases, the distribution approaches that of the GOE. Unlike the $HTT$ case however, now since two $\ket{HES}$ states are involved, the amplitude is a function of the scalar $\zeta_1 \cdot \zeta_2$. Interestingly, the amplitude with a non-zero $\zeta_1\cdot \zeta_2$, thus implying larger correlations among the in and the out states, show more RMT characteristics than with the $\zeta_1 \cdot \zeta_2 = 0$ case. This RMT description is for the non-coarse-grained 3 point amplitude. 
\item {\em Four point $HHTT$ amplitude chaos: } The 4 point function inherits the quantum chaotic behaviour almost trivially as it contains a {\em dressing} factor of the form of the 3 point amplitude which is chaotic. However, a priori, it isn't obvious if the chaos survives even in the probe limit. This is when in the $2\leftrightarrow 2$ scattering, the tachyon string comes in with extremely low energy. We focus in on this limit in \S\ref{subsec:probe} as this is the realm of classical chaotic scattering. We find that the leading contribution in this limit is given by a pole in the {\em s-channel}. The residue of the pole is microstate dependent, and once again shows features of level repulsion in its peaks distribution. 
\end{enumerate}

\item {\it How do the $HES$ amplitudes probe thermalization?} The notion of thermalization is emergent upon a suitable coarse-graining. In interacting quantum theories this notion is codified independent of any coarse-graining in the form of the Eigenstate Thermalization Hypothesis (ETH) \cite{Sred}. In its original form ETH is a statement about the matrix element of a local operator computed between two energy eigenstates having finite energy density : 
\begin{align}
\vev{ E_a | O | E_b } &= O_{th}(\bar{E} ) \delta_{a,b} + R_{ab} f_O(\omega) e^{-S_{\bar{E}}/2}. \label{eq:eth}
\end{align}
In the above expression $\bar{E} =  ( E_a + E_b)/2$ and $\omega = E_a - E_b$. Furthermore $R_{ab}$ is an antisymmetric random matrix, $f_O(\omega)$ is smooth function which contains information on thermal scales, and the suppressing factor $S_{\bar{E}}$ is the entropy computed at the average energy. The diagonal piece $O_{th}$ is the thermal expectation value of the observable corresponding to the temperature $\beta$ such that : $\bar{E} = \text{tr} \left( H e^{-\beta H} \right).$ ETH finds a natural place in the context of RMT \cite{ETH-Chaos}. The smooth function $f_O(\omega)$ as shown in \cite{MS}, from analyticity of the thermal Euclidean separated two point function, at intermediate $|\omega|$ decays exponentially,  bounded from below by $\exp\left( - \beta |\omega|/ 4 \right)$. Additionally, as $\omega \rightarrow 0$, $f_O(\omega)$ approaches a non-zero constant, and stays uniform at parametrically low frequencies \cite{Richter:2020bkf}. It is through these matrix element structures that an effective thermal description emerges for the eigenstates. 

\begin{enumerate}

\item In \S\ref{sec:HHT-average} we use the three point amplitude to build connections with the ETH. This is made possible since by the state-operator correspondence the vertex operators map to eigenstates of the worldsheet CFT. The $SL(2,R)$ invariance allows us to place the operators at 0, 1 and $\infty$. Therefore we are probing the matrix elements of the tachyon vertex operator in the massive DDF eigenstates. First we focus on the diagonal piece in Eq.\eqref{eq:eth}. This is effectively the thermal one point function. This allows us to draw a heuristic comparison with some recent results on $\vev{O_\Delta}_\beta$ in the $AdS/CFT$ context \cite{Grinberg:2020fdj} where the authors showed that under suitable analytic continuations the thermal 1-pt function captures the infalling time from the horizon to the singularity. We ask the question: {\it Under what coarse-graining precription, does the $\vev{H T H}$ amplitude capture the interior geometry of the black hole at the correspondence point?} Firstly we show analytically, at large $\ket{HES}$ level $N$, averaging the three point amplitude over all the states at the same level leads to exponentiation of the amplitude. Now, when we focus on the diagonal contribution, and further sum over all the scattering angles, we find that the amplitude is indeed consistent with what one will expect from the infalling time to the singularity for the black hole evaluated at the correspondence point. The analysis also shows that the other off-diagonal contributions are exponentially suppressed in the entropy, ${\cal{O}}\left( e^{- \sqrt{N}} \right)$ when $\ket{HES}$ is at level $N$.

\item {\it What does $\vev{HTH'}$ reveal with respect to the off-diagonal terms under coarse-graining?} In \S\ref{subsec:HHT-chaos-eth} we focus on the off-diagonal piece of Eq.\eqref{eq:eth}. We consider normalized $\ket{HES}$ states and evaluate the amplitude by summing over all degenerate states at different levels. This implements the {\em coarse-graining} whose absolute value we compare with ETH. We extract the factor $f_O(\omega)$ for different values of the scattering angle $\chi$, by taking into account the entropic suppression. This is numerically investigated for several cases, and we observe that $f_O(\omega)$ indeed approaches a finite value as $\omega \rightarrow 0$. For data obtained near $\chi \approx \pi/2$ we also find a region where $f_O(\omega)$ exponentially decays with $\omega$. 


\item {\it Is there a notion of typicality which arise in the $HES$ states? } Now that there is an evidence of the $\ket{HES}$ states being thermal, and since we are able to resolve microstates : which are the states at level $N$ that are most typical ? An answer to this question in the context of 2d CFTs have been answered in \cite{Datta:2019jeo}. The notion of typicality arises when the corresponding effective temperature is very high ($\beta \ll $ system size). As in the $\ket{HES}$ spectrum the states in a 2d CFT are arranged through partitions of integers, and hence for a given Verma module, can be described in terms of Young tableaux. The typical states are obtained from the typical Young tableaux diagram which is known to follow the Bose-Einstein distribution \cite{vershik1996statistical}. Therefore at high temperature one expects that partitioning into many low mode numbers to dominate compared to partitioning into fewer constituents of higher mode numbers. By analyzing numerically the averaged four point function in \S\ref{subsec:HHTT-numerics} we find that at large Mandelstam invariant $s$, the amplitude is dominated by low mode microstate. Hence if we associate higher energy scattering processes with typically higher effective temperatures, then the notion of {\it typicality} is consistent with the particular microstate dominance. 
\end{enumerate}

\item {\it What shape of the string does the four point function amplitudes reveal? } Given one has access to the 4 point function involving two $HES$ states, it is natural to ask what is the corresponding effective form factor? In context of Amati-Ruso like excited states, there is a precise answer to the form factor \cite{Manes:2003mw}. Here the excited state form factor calculation could be done analytically, and revealed a random walk description. The excited string, which is very long, at any instant resembles the worldline of a random walker \cite{Salomonson:1985eq, Mitchell:1987th}. The coarse-graining involves averaging over the intial states and summing over final states. We implement the same, while fixing the level $N$ of the $HES$ states and compute the form factor in \S\ref{sec:formfactor}. While our analysis is only semi-analytical, we do see in \S \ref{subsec:size} the characteristic $\vev{r^2} \sim L$ behaviour, which indicates random-walk. Here $\vev{r^2}$ is the mean square size of the long string whereas $L$ is its total length which goes like $\sqrt{N}$.

\end{enumerate}

\ndt While we make connections to black holes at the correspondence point using the S-matrix observables, there is an apparent puzzle with the worldsheet CFT spectrum which we know is integrable. Firstly it is known that in integrable systems ETH can still be valid {\em weakly} for typical states \cite{alba2015eigenstate}. This is to say that for coarse-graining inside a narrow energy window (which can be taken to approach zero in the thermodynamic limit) the variance of observables are polynomially suppressed in the system size instead of exponentially: as is the case in generic non-integrable systems. 

\ndt Excited states in integrable systems usually maybe approximable using Generalized Gibbs Ensembles, wherein one turns on chemical potentials corresponding to different integrable charges, which includes the Hamiltonian (see \cite{Banerjee:2019ilw} for free theory examples and \cite{Mandal:2015jla, Guo:2018pvi, Brehm:2019fyy} for 2D CFT discussions\footnote{Note, there exists exceptions to the generalized eigenstate thermalization hypothesis in integrable models as well : \cite{pozsgay2014failure}.}). Note that even within these examples there are cases where the GGE effectively becomes Gibbsian and therefore {\em thermality} emerges, see also \cite{Nandy:2016fwv}. It could be that the $\ket{HES}$ are part of this class of states. 

\ndt The exact mechanism however, is far from clear, heuristically : the amplitudes themselves being chaotic and exponentially many in number (in the fixed energy slice), when summed over and divided by total number of terms during coarse-graining, the denominator dominates and exponentially so. There are many cancellations in the numerator due to the chaotic nature of the amplitudes. This results in the consistency with the expectation of ETH with the emergent {\em thermality}, making features consistent with the black hole at the correspondence point. In the $AdS/CFT$ context some interesting results of smooth horizon physics emerging at large $N$ starting from microstates can be found in \cite{Burman:2023kko}. Even in \cite{Burman:2023kko} at finite $N$, coarse-graining at fixed energy slice plays a key role. 

\ndt In addition to the above features, one should be aware that the black hole geometry imposes unique constraints on the S-matrix \cite{Polchinski}. In particular interchanging the incoming and outgoing quanta relate the two different S-matrices by a factor coming from the time-delay of the corresponding modes. It will be interesting to explore this effect in a higher point generalization of the kinds of amplitudes we study involving $\ket{HES}$ states.  As a step towards this and .... we sketch an algorithm to obtain amplitudes of the form $\vev{ H_1 H_2  T_1 T_2 \cdots T_n }$ in Appendix \S\ref{app:higher}. We end by mentioning some more future directions in \S\ref{sec:future}. 

\subsection*{Online Resource}
\ndt We have made the \texttt{Mathematica} notebooks available for public use. The notebooks can be used for direct evaluation of the $HES$ amplitudes as well as for doing statistics using peak positions and amplitudes. The files are available at the \texttt{github} page of one of the authors : \url{https://github.com/anurag-dot-physx/HES-S-Matrix}

\section{Computation of $HES_1(p_1,\zeta_1) \rightarrow HES_2(p_2, \zeta_2) + T_3(p_3)$}
\label{sec:HHT}
\begin{figure}
\centering
\rotatebox{0}{\includegraphics*[width= 0.5 \linewidth]{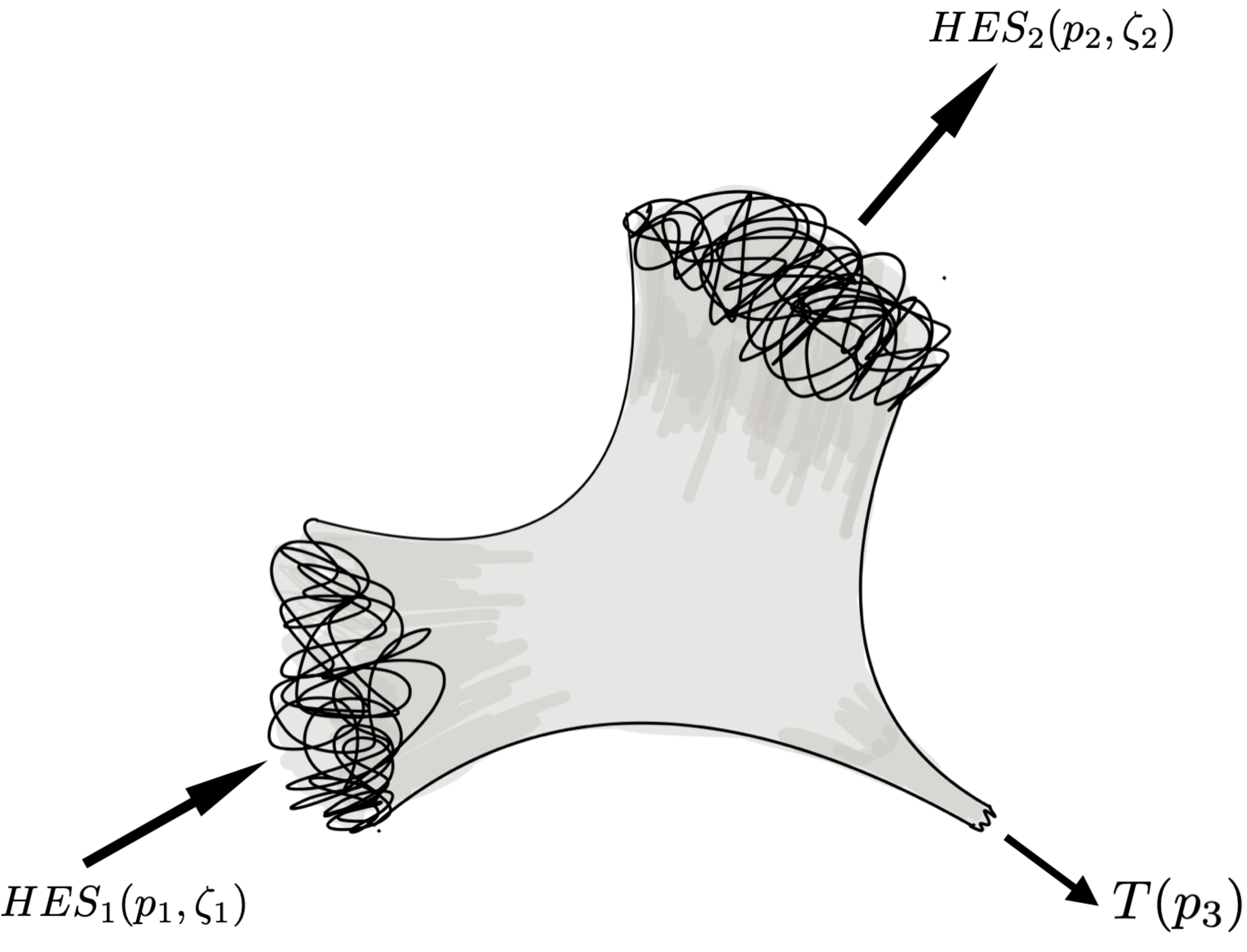}}
\rotatebox{0}{\includegraphics*[width= 0.42 \linewidth]{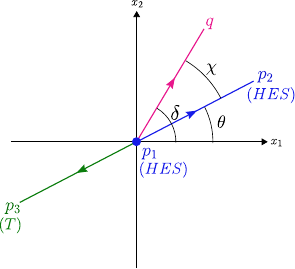}}
\caption{Left Panel : The 3 point scattering set-up considered. Right Panel : Kinematics with angles indicated, green arrows indicate tachyons, while blue the $HES$ ones. $HES_1$ is initially at rest, the pink arrow indicates the photon momenta which make up the $\ket{HES}$ states.}\label{fig0}
\end{figure}

In this section we consider the amplitude for the process of a $\ket{HES}$ decaying into another $\ket{HES}$ along with a $Tachyon$. The $\ket{HES}$ that we consider here are reviewed in detail in \cite{GR} where we refer the reader to. The quantum numbers of the $\ket{HES}$ microstates consists of its target space momenta $p =  \tilde{p} - N q$, where $\tilde{p}$ and $q$ are the momenta of the tachyon and the photons that take part in the DDF construction. There are $N$ photons taking part in the construction, which raise the energy of $\ket{HES}$. Additionally, there is also the polarization $\zeta$ which is given in terms of the constitutent photon polarization $\lambda$ by: $\zeta = \lambda - (\lambda \cdot p ) q$. For tractability we choose all the photons parallel and all of them to have the same polarization. We have $\zeta \cdot p  =  0$ and the mass-shell condition $M^2 = (N-1)/\alpha'$ which fixes $\tilde{p}\cdot q = p \cdot q = 1$. To simplify contraction we also chose $\lambda^2 = 0$. In terms of DDF creation operators, $A_{-r}$, the state is given by : 
\begin{equation}
\prod_{r} \left( \lambda \cdot A_{-r} \right)\ket{0, \tilde{p}} 
\end{equation}
Note that we have admitted repeatition of indices at this stage. Also note that the state is not normalized. For the ETH criteria we need to consider normalized microstates which is discussed in \S\ref{app:norm}. Carrying out contraction lead to:
\begin{equation}
: \lambda \cdot A_{-r} e^{i \tilde{p} \cdot X } : = \sum_{m=1}^r\frac{i}{(m-1)! } \zeta \cdot \partial^m X\,\, S_{r-m} \left( -\frac{i r}{s!} q\cdot \partial^s X \right) e^{i p \cdot X },
\end{equation}
where $S_a(b_s) = S_a(b_1, b_2 \dots, b_a) $ denotes a Schur polynomial. We have chosen $\alpha' = 1/2$. To make notations convenient let us introduce : $X_i = X(z_i)$ and $S_{r-n}^i = S_{r-n}\left( - \frac{ir}{s ! } q_i \cdot \partial^{s} X_i \right)$. The $z_i$'s denote the disk coordinates where the three vertex operators are inserted. These are on boundary of the disk and hence are mapped to the real axis, thus $z_i = \bar{z}_i$. Now we evaluate the amplitude as in Fig.\ref{fig0}. 

\ndt The final amplitude is obtained after two steps. Firstly we need to carry out contractions of the relevant string vertex operators, next we need to integrate over the vertex coordinates. The integrand that we need to compute is given by: 
\begin{align}
e^{\cal L} &= \langle :\prod_{r_1} \left( \sum_{m_1}^{r_1} \frac{i }{(m_1 - 1)!} \zeta_1 \cdot \partial^{m_1} X_1 S^1_{r_1-m_1} \right) e^{i p_1 \cdot X_1}:: \prod_{r_2} \left( \sum_{m_2}^{r_2} \frac{i }{(m_2 - 1)!} \zeta_2 \cdot \partial^{m_2} X_2 S^2_{r_2-m_2} \right) e^{i p_2 \cdot X_2}:\nonumber \\
&: e^{i p_3 \cdot X_3 }: \rangle \label{eq:uncontract}
\end{align}
We evaluate the contractions using the convention $\langle X^{\mu}(z)X^{\nu}(w) \rangle =-\eta ^{\mu \nu}\log(z-w)$, which results in: 
	\begin{equation}
		\begin{split}
			\sum_{k=0}^{\lbrace J_1,J_2 \rbrace_{min}} \sum_{\substack{\lbrace a(k)\rbrace \subset \lbrace r_{1} \rbrace\\ \lbrace b{(k)}\rbrace \subset \lbrace r_{2} \rbrace\\ \text{ordered}}} \frac{1}{k!}&\prod_{i=1}^{k} \Bigg[ \sum_{m_1,m_2=1}^{a_i, b_i} \bigg \lbrace \frac{(-1)^{m_2+1}(m_1+m_2-1)! \zeta_1.\zeta_2}{(m_1-1)!(m_2-1)! z_{21}^{m_1+m_2}}  \bigg \rbrace \\
			& \;\ \;\ \;\ \;\ \;\ \;\ \times S_{a_i-m_1} \bigg( \frac{a_i}{s_1} \left(  \frac{q_1 \cdot p_2}{z_{21}^{s_1} }+ \frac{q_1 \cdot p_3}{z_{31}^{s_1} } \right) \bigg) S_{b_i-m_2} \bigg( \frac{b_i}{s_2} \left(  \frac{q_2 \cdot p_1}{z_{12}^{s_2} }+ \frac{q_2 \cdot p_3}{z_{32}^{s_2} } \right) \bigg) \Bigg]\\
			\times &\prod_{l_1 \in \overline{\lbrace a(k) \rbrace}} \Bigg[ \sum_{n_1=1}^{l_1}\bigg(\frac{ \zeta_1 \cdot p_2 }{z_{21}^{n_1} } +  \frac{ \zeta_1 \cdot p_3 }{z_{31}^{n_1} }  \bigg)   S_{l_1-n_1} \Bigg]\\ 
			\times 
			&\prod_{l_2 \in \overline{\lbrace b(k) \rbrace}} \Bigg[ \sum_{n_2=1}^{l_2}\bigg(\frac{ \zeta_2 \cdot p_1 }{z_{12}^{n_2} } +  \frac{ \zeta_2 \cdot p_3 }{z_{32}^{n_2} }  \bigg) S_{l_2-n_2} \Bigg] \times \prod_{i<j} z_{ij}^{p_i \cdot p_j } = B_{\{r_1\},\{r_2\}}\prod_{i<j} z_{ij}^{p_i \cdot p_j } .
		\end{split}
		\label{ampHHT}%
	\end{equation}
	
\ndt The first term in the square bracket includes terms coming from contractions among the DDF vertex operators. The sum over $\lbrace a(k)\rbrace$ and, $\lbrace b(k)\rbrace$ denote the different ordered subsets of partitions of $N_1$ and $N_2$, which are $\lbrace r_{1} \rbrace$ and, $\lbrace r_{2} \rbrace$ respectively. They have $k$ elements each, which get contracted. Taking ordered subset is important because we want to include all possible $k$-term contractions for different values of $k$. For example if $k=2$ then $\lbrace a(k)\rbrace_1=\lbrace a_1,a_2\rbrace \;\ ; \;\ \lbrace a(k)\rbrace_2=\lbrace a_2,a_1\rbrace \;\ ; \;\ \lbrace b(k)\rbrace_1=\lbrace b_1,b_2\rbrace \;\ ; \;\ \lbrace b(k)\rbrace_2=\lbrace b_2,b_1\rbrace$ must be considered in order to consider all possible contractions i.e. $p(a_1,b_1) \times p(a_2,b_2)$ and $p(a_1,b_2)\times p(a_2,b_1)$. We have to divide by ${k!}$ to compensate for the overcounting.  

\ndt The number of such contractions depends on the number of $A_{-r}$'s present in a DDF factor. This is given by $ J = \sum_{r} n_r$, where $n_r$ counts the repeatitions of the mode $r$ in the partition $\{r \}$. Since there are two DDF factors the maximum contractions $k$, is given by the minimum among $J_1$ and $J_2$. The complementary subsets are the remaining operators that get contracted amongst each other to produce the products of the second and the third square brackets of Eq.\eqref{ampHHT}. 

\ndt Finally, in writing down the result we also make the choice of $q_1 \propto q_2$. This gets rid of contractions between the derivatives and the arguments of the Schur polynomials.

\subsection{Kinematics for $HHT$ }	
We follow the metric signature $(-,+,+\dots +)$. Therefore the on-shell condition is $p_i^2 = - M_i^2$. Furthermore we take all the momenta to be ingoing. We choose the frame such that $\ket{HES_1}$ at level $N_1$ with rest mass $\sqrt{2(N_1-1)}$ decays into a tachyon of rest mass squared $-2$ and $\ket{HES_2}$ at level $N_2$ with rest mass $\sqrt{2(N_2-1)}$. As shown in Fig. \ref{fig0} we take $\ket{HES_1}$ at rest. Explicit parametrizations of the momenta, as well as that of the DDF polarizations are presented in Appendix \S\ref{app:kin}. For simplicity we take the $HES$ photon momenta to be proportional to each other with $q_2 = -\gamma q_1$. We will consider and contrast two choices of polarizations : $\zeta_1\cdot\zeta_2 = 1$ and when it is zero in what follows.  

\subsection{Evaluating the $HHT$ amplitude}
The tachyon contraction factor of Eq.\eqref{ampHHT} can be explicitly evaluated using the constraint relations in \eqref{constraint}. 
%
We replace $p_3$ in our computation as it turns out that the final relations are relatively simpler compared to any other substitutions. This gives: 
\begin{equation}
 \int \prod_{i} d z_i \;\ e^{\mathcal{L}} = \int \frac{d z_1 d z_2 d z_3}{z_{12} z_{13} z_{23}} 
			\exp \left( \log \left( \frac{z_{12}^{N_1+N_2}}{z_{13}^{N_2-N_1} 
				z_{23}^{N_1-N_2}}\right) +   \log B_{\{r_1\},\{r_2\}} \right).
		\label{sl2}
\end{equation}
	The quantity in front of the exponential with the measure becomes $SL_2$ invariant. 
	Thus by using $SL_2$ invariance and fixing, $z_1 = 0, z_2 = 1$ and $z_3 =\infty$ (which is decided by the first term in the exponent), we can get rid of the integrals completely. Finally, the total amplitude becomes:
	\begin{equation}  \label{HHT amp}
		\boxed{
			\mathcal{A}=\sum_{k=0}^{min \lbrace J_1, J_2 \rbrace}\frac{1}{k!}  \sum_{\substack{\lbrace a(k) \rbrace \\ \lbrace b(k) \rbrace  \\ \text{ordered}} } \left(\prod_{i=1}^{k} P(a_i,b_i) \prod_{l_1 \in \overline{\lbrace a(k) \rbrace}}Q_1(l_1) \prod_{l_2 \in \overline{\lbrace b(k) \rbrace}}Q_2(l_2) \right)
		},
	\end{equation}
where, the factors $P(a_i,b_i) ,~ Q_1(l_1) ~ \text{and} ~ Q_2(l_2)$ are written out explicitly in Appendix \S\ref{app:PQ}. 
Our final 3 point amplitude is a function of the angular difference between the DDF photon and the $\ket{HES_2}$ momenta directions, which we call  $\chi$ as shown in Fig. \ref{fig0} . 
%
%

\section{Averaging the $HHT$ scattering amplitude} 
\label{sec:HHT-average}
In this section we average the amplitude over $HES$ states belonging to the same level (mass). This translates to averaging over different partitions. This averaging takes the form: 
	\begin{align}
		\vev{\mathcal{A}_{(\lbrace r_1 \rbrace,\lbrace r_2 \rbrace)}}_{F}  \sim (-1)^{N}    \frac{1}{\Omega (N_1)\Omega(N_2)} \sum_{ \{n_1\}, \{n_2\} }  \mcA \left(  \{n_1\}, \{n_2\}  \right),
		\label{eq:HHTfull}
	\end{align} 
where $\Omega(N)$ counts the number of $\ket{HES}$ microstates at the fixed level $N$, as discussed in Appendix \S\ref{app:omega}. At large $N$, we are able to analytically carry out averaging in the kinematic choice where $\zeta_1 \cdot \zeta_2=0$. Now the amplitude is of the product form $$\mathcal{A}_{(\lbrace r_1 \rbrace,\lbrace r_2 \rbrace,\zeta_1 \cdot \zeta_2=0 )}= \prod_{r_1}\left(\zeta_1 \cdot p_2V_1(r_1)\right) ^{n_{r_1}}\prod_{r_2}\left(\zeta_2 \cdot p_1V_2(r_2)\right)^{n_{r_2}},$$ where the exact form of the $V_1$ and $V_2$ are written down in Appendix \S\ref{app:PQ}. Therefore the full averaged amplitude becomes : 
	\begin{align} 
		\vev{{\cal A}}_F &=  \sum_{\substack{\lbrace n_{r_1} \rbrace \\
				\lbrace n_{r_2} \rbrace}}\frac{1}{\Omega(N_1) \Omega(N_2) } \oint d\beta_1 d\beta_2 \,\, e^{ \beta_1 N_1 +\beta_2 N_2 }e^{- \beta_1 \sum_{r_1} r_1 n_{r_1} }e^{- \beta_2 \sum_{r_2} r_2 n_{r_2} } \mathcal{A}_{(\lbrace r_1 \rbrace,\lbrace r_2 \rbrace,\zeta_1 \cdot \zeta_2=0 )}\nn  \\
		&=  \frac{1}{\Omega(N_1) \Omega(N_2) } \oint d\beta_1 d\beta_2 \,\, e^{ \beta_i N_i } \prod\limits_{r_1 } \bigg( 1 - e^{-\beta_1 r_1}(\zeta_1 \cdot p_2) V_1 \bigg)^{-1} \prod\limits_{r_2 } \bigg( 1 - e^{-\beta_2 r_2} (\zeta_2 \cdot p_1) V_2 \bigg)^{-1}. \label{eq:HHTfull}
	\end{align}
We evaluate the above expression for $N_1=N_2=N$ and $N \rightarrow \infty$ where both $V_1$ and $V_2$ in the above expression becomes unity and $(\zeta_1 \cdot p_2) \sim -\sin \chi, \,\,\,(\zeta_2 \cdot p_1) \sim -\sin \chi$. Plugging these into Eq.\eqref{eq:HHTfull} we obtain: 	
		\begin{align}
			\langle \mathcal{A}\rangle_F &\simeq \frac{(-1)^N}{\Omega_1(N)\Omega_2(N)} \oint d \beta_1 e^{\beta_1 N} \prod_{r_1}\left( 1+\sin \chi (e^{-\beta_1})^{r_1}\right)^{-1} \oint d \beta_2 e^{\beta_2 N} \prod_{r_2}\left( 1+\sin \chi (e^{-\beta_2})^{r_2}\right)^{-1}.
		\label{ampav}
	\end{align} 
At large $N$ we expect the saddle point of $\beta_i$ integrals to be dominated by small values. It is under this expectation that the products maybe evaluated in terms of dilogarithm function: 	
	\begin{align}
			\prod_{r_1}\left( 1+ e^{-\beta_1 r_1 } \sin \chi\right)^{-1}&=\exp \left[\sum_{k=1}^{\infty} \frac{(-\sin \chi)^k}{k \big(e^{\beta_1 k}-1\big)}\right] 
			\simeq \exp \left[ \sum_{k=1}^{\infty} \frac{(-\sin \chi)^k}{k^2 \beta_1}\right] 
			 =\exp \left[ \frac{1}{\beta_1} \text{Li}_{2}(-\sin \chi) \right]. \label{eq:dilog}
		\end{align}
 The saddle of the $\beta_1$ and $\beta_2$ integrals are at, 
 $
 \beta_i^*  = \sqrt{\frac{1}{N} \text{Li}_{2}(-\sin \chi)}. 
 $
 Hence each of the on-shell integrals evaluates to:
	\begin{align}			
	\oint d \beta e^{\beta N} \prod_{r}\left( 1+ e^{-\beta r } \sin \chi \right)^{-1}
			&\simeq \exp \left[2\sqrt{N} \sqrt{\text{Li}_{2}(-\sin \chi)} \right].
		\end{align}  
Collecting all the factors we obtain: 	
	\begin{align}
			\langle \mathcal{A}\rangle_F & \simeq (-1)^N\exp \Bigg[4\sqrt{N}\bigg( \sqrt{\text{Li}_{2}(-\sin \chi)}-\frac{\pi}{\sqrt{6}}\bigg) \Bigg] .
			\label{hhtaverage}
	\end{align}
	
\subsubsection*{The diagonal piece at same level}
Out of the summation we can separate out the diagonal piece, i.e., the contribution coming from terms where both the in and the out states are described by identical partitions. Following the above methods we find: 
	\begin{equation}
			\boxed{ \vev{{\cal A}}_D = (-1)^N \exp \left[2\sqrt{N}\bigg\lbrace \sqrt{\text{Li}_{2}(\sin^2 \chi)}-\frac{\pi}{\sqrt{6}}\bigg\rbrace \right]} .
	\label{eq:HHTDiag}	
	\end{equation}

\subsection*{Heuristic comparison with the thermal one-point function} 

\begin{figure} 
\centering
\includegraphics*[width= 0.5 \linewidth]{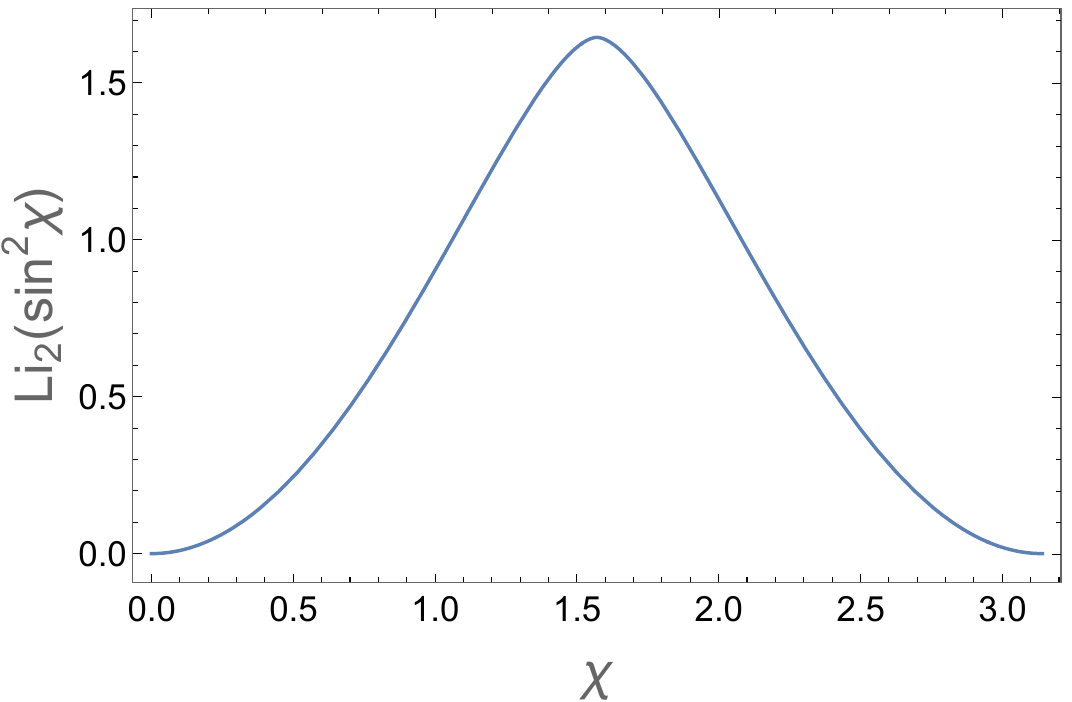}
\caption{$\chi$ dependent part of $\log \vev{{\cal{A}}}_D$ showing a maxima at $\chi = \pi/2$.}\label{figLi}
\end{figure}

To compare with the thermal one point function we need to consider only the diagonal matrix elements, i.e., $\ket{HES}$ states having identical partitions. Therefore the relevant expression is the coarse-grained expression as given above in Eq.\eqref{eq:HHTDiag}. 
As shown in Fig. \ref{figLi}, the function $\sqrt{\text{Li}_2 (\sin^2\chi )}$. 
\ndt The value at the maximum of Fig.\ref{figLi} is $\text{Li}_2 (1 ) = \frac{\pi^2}{6}$. Note, that the square root of this precisely cancels the $\pi/\sqrt{6}$ factor in the exponent of Eq.\eqref{eq:HHTDiag}. Hence if we sum over all the angles, then upto exponentially small corrections : 
\begin{equation}
\vev{A}_D \sim 1 + {\cal{O}}\left( e^{- \sqrt{N} } \right) , \label{eq:ad}
\end{equation}
It is known that in semi-classical scalar absorption in case of rotating black holes, the scattering cross-section increases as the incident probe scalar deviates from on-axis orientation and as the black hole angular momentum decreases \cite{Macedo:2013afa}.  Our observation of maximal amplitude at $\chi = \pi/2$ is consistent with these.

\ndt In analogy with ETH, Eq. \eqref{eq:eth}, we will like to identify this diagonal piece with the thermal one point function. And, since this is the one-point function computed in the coarse-grained $\ket{HES}$ background, we will compare this with the black hole one point function, where the black hole is at the correspondence point. In the $AdS/CFT$ context, \cite{Grinberg:2020fdj} pointed out that the CFT thermal one point function, under suitable analytic continuation of the conformal dimension of the operator, contains a phase which probes the geometry behind the horizon: $\vev{ O }_\beta \sim \exp\left( -i \, m\, t_s  \right)$, where $t_s$ is the time to singularity in an asymptotically $AdS$ black hole. This identification has been shown to hold rigorously for many generalizations \cite{David:2022nfn}.  As the following comparison is at the level of order of magnitudes, for simplicity we consider the non-rotating case. We take the correspondence point black hole to be described by the Schwarzchild  metric, which in $d=4$ is : 
\begin{align}
ds^2 &= - \left( 1- \frac{2G M}{r} \right) dt^2 + \left( 1- \frac{2G M}{r} \right)^{-1}dr^2 + r^2 d\Omega^2.
\end{align}	
It turns out that the time to the singularity for the above metric yields a finite quantity : 
\begin{align}
t_s &=\lim_{\epsilon\rightarrow 0 } \int_{2 G M}^\epsilon \frac{dr}{ -\sqrt{  \frac{2G M}{r } -1 } } 
=  G \, M \pi + {\cal{O} }\left( \epsilon^{3/2} \right).
\end{align}
Note, unlike $AdS$ the $t_s$ depends explicitly on the black hole mass, also note that we work with the negative square-root to get a positive physical infall time. Hence we expect that: 
\begin{align}
\vev{O}_\beta &=  \exp\bigg[   \frac{\pi \, G \, M}{\sqrt{\alpha'}}    \bigg] =   \exp\bigg[  \pi \sqrt{\alpha'} g_s^2 \, M  \bigg].
\end{align}
In the above we used the tachyon mass $ = i/\sqrt{\alpha'}$ and the gravitational coupling $G = g_s^2 \alpha'$. Now at the correspondence point $g_s^2 M = 1/\sqrt{\alpha'}$, hence we are simply left with an ${\cal{ O}}(1)$ real number $\sim e^\pi$. In this sense it is consistent with the ${\cal{O}}(1)$ number obtained in Eq. \eqref{eq:ad}.

\subsubsection*{The off-diagonal piece at same level}
We can now write the off-diagonal averaged amplitude in terms of Eq.\eqref{eq:HHTfull} and Eq.\eqref{eq:HHTDiag} :
	\begin{align}
		\vev{\mcA}_{OD}  &\simeq  \frac{1}{\Omega(N)^2}  \left(  \sum_{ \{n_1\}, \{n_2\} }  \mcA \left(  \{n_1\}, \{n_2\}  \right)  -  \sum_{ \{n\} }  \mcA \left(  \{n\}, \{n\}  \right)  \right)  \equiv  (-1)^{N} \left(  u e^{i \phi} - v \right). \nn
	\end{align}
where we have defined real quantities $u, v$ and $\phi$ through: 
	\begin{align}
		u \equiv &\exp \left( \frac{- 4 \pi}{\sqrt{6}} \sqrt{N} \right), ~  v \equiv  \exp \left[  2 \sqrt{N} \left\{  \sqrt{\text{Li}_2 ( \sin^2 \chi )} - \frac{2 \pi}{\sqrt{6}}  \right\}  \right] , ~ e^{i \phi}  \equiv  \exp \left[  4 \sqrt{N} \sqrt{\text{Li}_2 ( - \sin \chi )}  \right] .\nn
	\end{align}
Therefore the absolute value squared which is $|\vev{\mcA}_{OD}|^2  = u^2 + v^2 - 2 u v \cos \phi $ is bounded by: 
	$$  (u-v)^2  \leq |\vev{\mcA}_{OD}|^2  \leq  (u + v)^2.
$$	
At large $N$ the upper bound behaves as, $ \sim  \exp \left[  -2 \sqrt{N} \left( \frac{2 \pi}{\sqrt{6}}  -  \sqrt{\text{Li}_2 ( \sin^2 \chi )} \right)   \right]$. Since ${\text{Li}_2 ( 1 )}= \pi/ \sqrt{6}$, which is its maximum value, we obtain:
	\begin{align}
		|\vev{\mcA}_{OD}| \leq  \exp \left( - \frac{2 \pi}{\sqrt{6}} \sqrt{N}  \right)  \sim   \left(\Omega (N)  \right)^{-1} .
	\end{align}	
Hence, when the microstates are not the same, there is entropic suppression. Next, we will numerically probe off-diagonal elements at different levels. 

\subsection{Numerically probing chaos and the ETH} 
\label{subsec:HHT-chaos-eth}
Unlike the diagonal amplitude, the off-daigonal amplitude being not amenable to analytic averaging at large $N$, we investigate its features numerically. 
\subsubsection{Signatures of RMT chaos in the $HHT$ amplitude}
\begin{figure} 
\centering
\rotatebox{0}{\includegraphics*[width= 0.48 \linewidth]{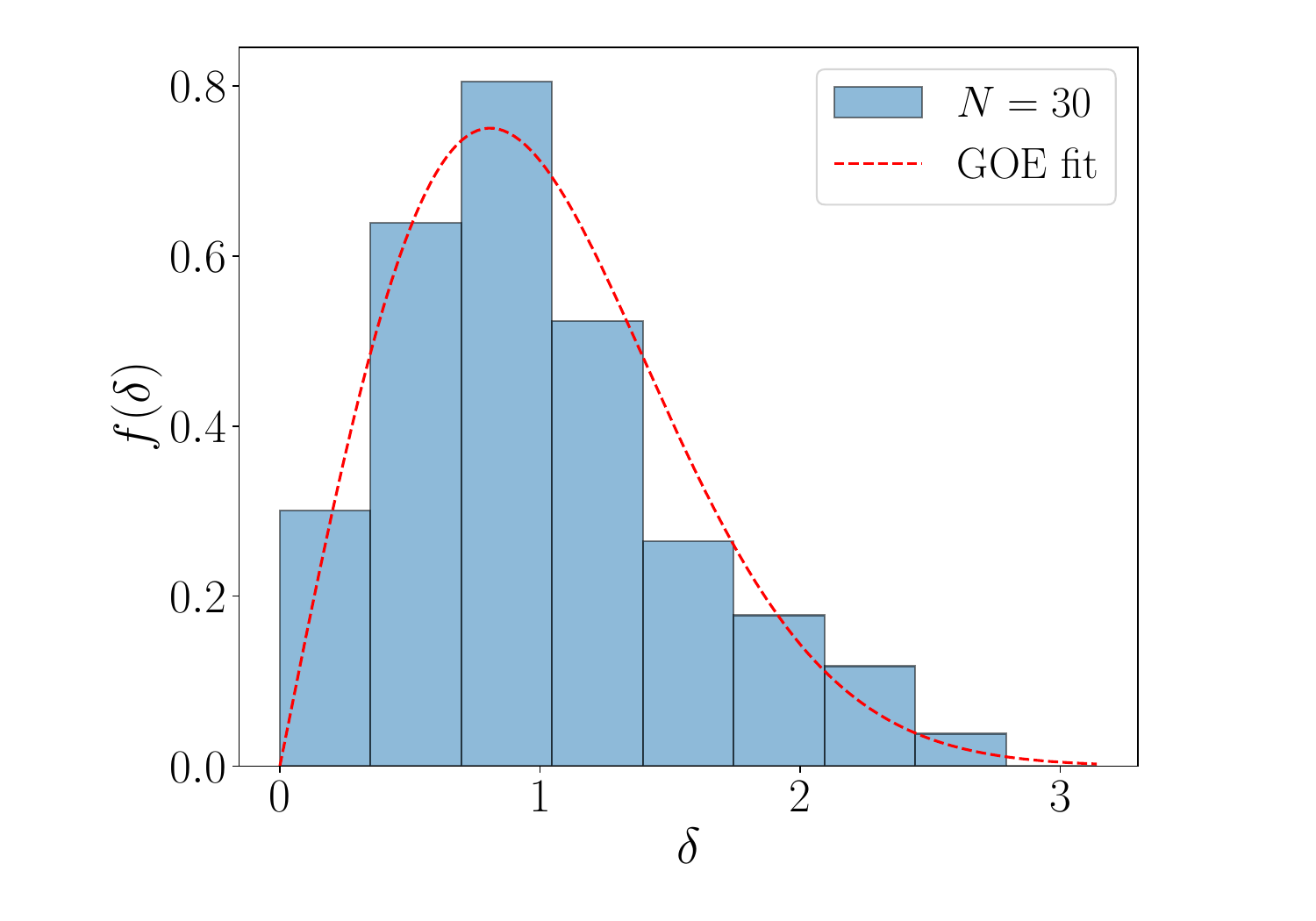}}
\rotatebox{0}{\includegraphics*[width= 0.48 \linewidth]{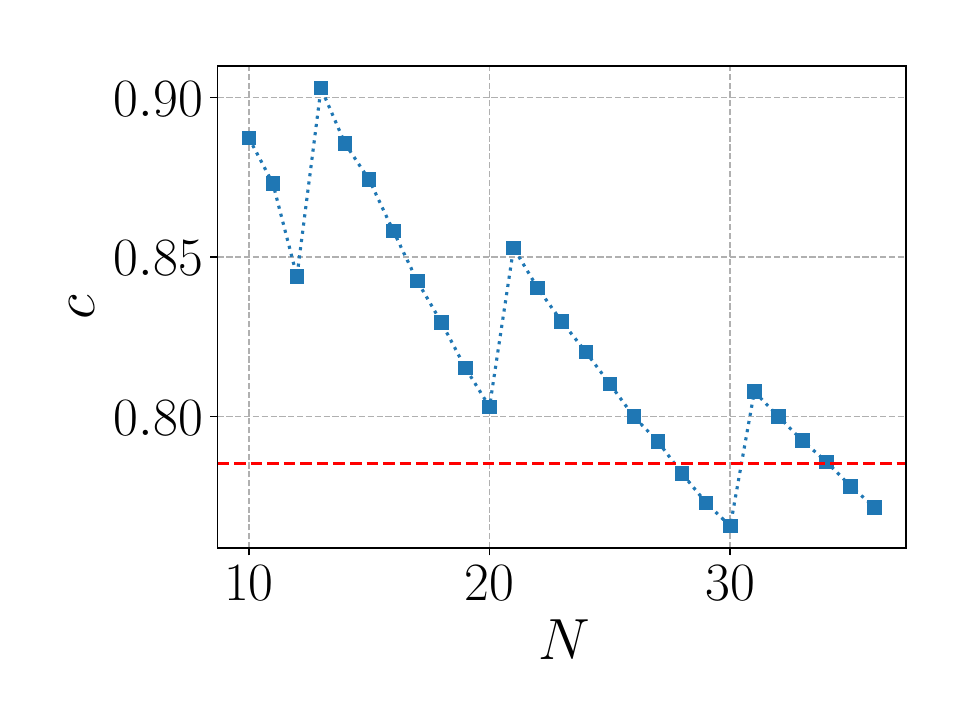}}
\caption{Left Panel : Plot of the peak level spacing statistics of the amplitude with $\zeta_1 \cdot \zeta_2 = 0$ and identitcal partitions for $HES$ at $N=30$. The total number of peaks for this plot is $25664$. The red line shows fit to $p_{}(s)$ with fit parameter $c= 0.7656$.  Right Panel : We plot the fitted $c$ parameter as a function of $N$ from 10 to 36. The dashed red line indicates the GOE value $c = \pi/4$. }\label{Fig2}
\end{figure}

As indicated in \cite{GR, BF2} we focus on the resonance peak statistics to probe chaos. We consider the $HHT$ amplitude for all the partitions of $N_1=N_2=N$. 
%
The steps of the numerics are following:
	\begin{enumerate}
		\item For two fixed partitions $N$ : $\{r_1\},\{ r_2\}$ amplitude $\mathcal{A}(\chi)$ is evaluated in the range of $\chi= \left[ 0, \pi \right]$ with step size of $\delta \chi = 0.001$.
		\item From the amplitude array, the peak positions of $\mathcal{A} (\chi)$ are found out by checking which $\{\chi_*\}_{\{r_1\},\{r_2\}}$ values satisfy $\mathcal{A}' (\chi_*) = 0$.
		\item We repeat the above two steps for all possible sets $\{r_1\}, \{r_2\}$ and keep adding to : $\{ \chi_* \}$.  Next, the probability distribution of these peak positions and the spacings between the adjacent peaks are studied.
	\end{enumerate}
To look into the chaotic features of the amplitude, we look at the level statistics and find numerical evidence of level repulsion. See left panel of Fig. \ref{Fig2} where we present a histogram for the $\zeta_1 \cdot \zeta_2 = 0$ case along with the fitted Wigner-Dyson distribution corresponding to GOE. In order to reach larger $N$ values we have considered identical partitions for the $HES$ states. In the right panel of Fig.\ref{Fig2} we show an indicator for comparing the level statistics of the amplitude peaks with that of GOE random matrix. We have fitted the unfolded data\footnote{ The details of unfolding is mentioned in Appendix \S \ref{app:unfolding}. } with the distribution : $p_{}(s) \propto  s  \exp\left( -c \,s^2 \right)$, the GOE value being $c = \tfrac{\pi}{4} = 0.785298$. We observe that as $N$ increases the $c$ parameter comes closer to the GOE value and keeps oscillating around it. To get an idea of the number of data points used for the plot : for $N=36$ the total number of partitions is $17977$ and with our resolution of $\Delta \chi = 0.001$ we find 82263 peaks.

\ndt The level repulsion was also noted for the $HTT$ amplitude in \cite{BF2, Bianchi:2024fsi}, hence is not surprizing that it continues to hold for the $HHT$ case. In the $HTT$ case when unfolded, the authors observed fit to the GUE random matrix ensemble, whereas for $HHT$ the fit is with GOE. We ascribe this difference to the fact that the amplitude in our set-up with identical $HES$ states in the initial and final configurations is invariant under time-reversal \cite{dyson1962threefold}.

\begin{figure} 
	\centering
	\rotatebox{0}{\includegraphics*[width= 0.48 \linewidth]{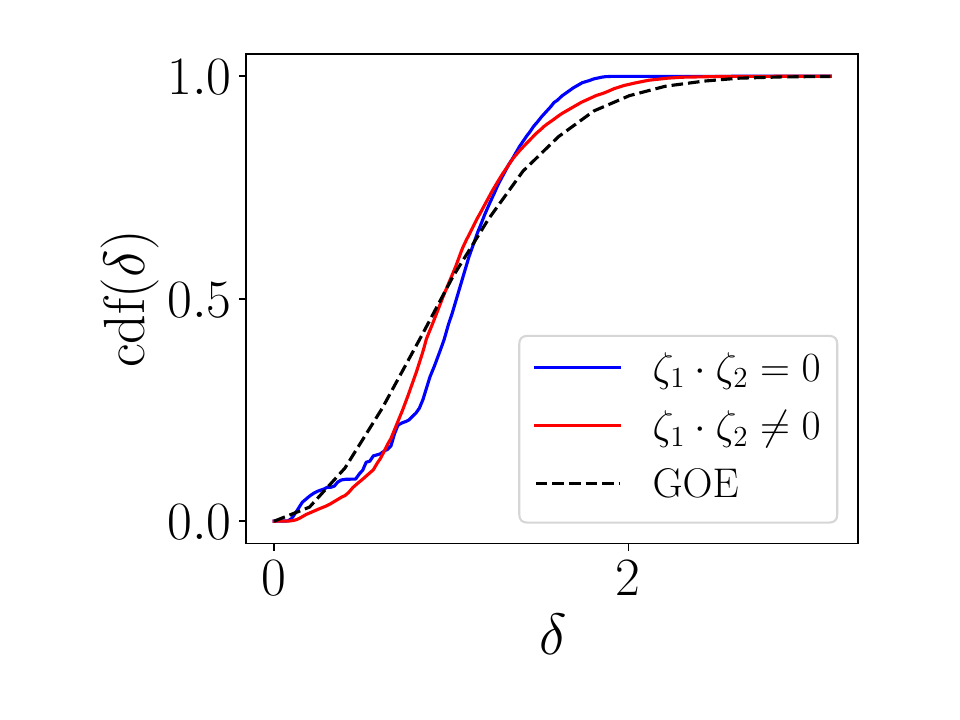}}
	\rotatebox{0}{\includegraphics*[width= 0.48 \linewidth]{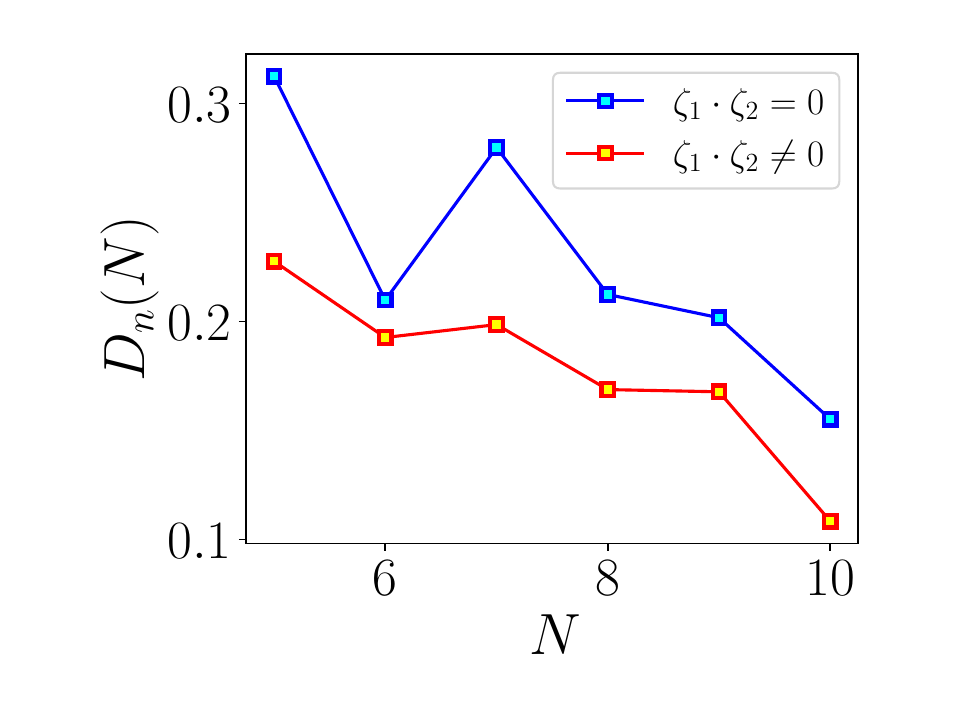}}
	\caption{Left Panel : Plot of the peak level spacing statistics CDF of the amplitude with $\zeta_1 \cdot \zeta_2 = 0$, $\zeta_1 \cdot \zeta_2 \neq 0$ and GOE random matrix statistics. Right Panel : KS statistical distance between the level spacing CDF of the amplitude and GOE, showing that $\zeta_1 \cdot \zeta_2 \neq 0$ case is closer to GOE for same values of $N$.}\label{Fig3}
\end{figure}

\ndt Next we look at the effect of the polarization in chaos. We therefore consider both choices of polarizations, such that $\zeta_1 \cdot \zeta_2 = 1$ and $0$. We use the Kolmogorov-Smirnov (KS) test \cite{ref1}. This compares the empirical cumulative distribution (ECDF) functions of two data sets and provides a measure to decide how close the two distributions are.  The ECDF of a given data set $\{x_1, \cdots ,x_n\}$ is:  $F_{n} (x) = \tfrac{\text{number of elements $\{x_i\} \leq x$}}{n}$
The KS statistic for a given CDF $F(x)$ is defined via: $D_{n} = \sup_{x}  |F_{n}(x) - F(x)|$. Thus smaller $D_{n}$ indicates that the two distributions are closer. Here, we take $F(x)$ from GOE Wigner-Dyson distribution and compare it with the ECDF derived from the peak statistics. We perform the KS test for same values of $N$ for the two cases of $\zeta_1 \cdot \zeta_2$, see Fig.\ref{Fig3}. We find that a non-zero inner product between the polarizations approach the chaotic distribution for a smaller $N$ compared to the orthogonally polarized case. A likely explanation for this difference is the fact that $\zeta_1\cdot \zeta_2 =0$ is a more restrictive kinematic choice than $\zeta_1 \cdot \zeta_2\neq 0$. In particular, for the latter the phase space volume being more, notions of thermalization are satisfied at a smaller $N$ value. \footnote{We thank Arnab Sen for pointing this out to us.} 

\subsubsection{Probing off-diagonal matrix elements in the $HHT$ amplitude}
We consider coarse-graining over the two $HES$ states in $\vev{ HES_1 (p_1, \zeta_1) T(p_3) HES_2 (p_2 , \zeta_2 )}$. The $\ket{HES}$ states are now taken to be at different levels, neverthless we average over all states at a given level. The information that we infer from this amplitude is the same present in the off-diagonal piece of Eq.\eqref{eq:eth}. Fixing off-diagonal row \& coloumn, i.e., fixed $N_i$ and $N_j$, we average over the amplitudes (matrix elements) which relate these two energy levels:
\begin{align}
|{\vev{\tilde{A}}}_{ij} |^\chi &= \bigg(  \frac{1}{\Omega(N_i) \Omega(N_j)}\,\,  \bigg|  \sum_{\substack{ H_i \in \{ N_i \}   \\  H_j \in \{ N_j \}  } }  \frac{\vev{ H_i T H_j   }^\chi}{\sqrt{\vev{ H_i | H_i }}\sqrt{\vev{ H_j | H_j }}} \bigg| \bigg), \quad i \neq j.\label{eq:tilde}
\end{align}
Note, that after summing over states with fixed $N_i$ and $N_j$, we have taken an absolute value, hence we compare this with $| e^{- S(\bar{E})/2} f(\omega) R_{ij} | = e^{- S(\bar{E})/2} f(\omega) | R_{ij} |$, where $S(\bar{E})$ denotes the entropy corresponding to the states at average energy between $E_i$ and $E_j$. It is interesting to note that this is the same coarse-graining as in the case for probing ETH in $c>1$ CFTs for heavy states \cite{Brehm:2018ipf, Romero-Bermudez:2018dim, Hikida:2018khg}, where the analytical form of $f_O(\omega)$ revealed the quasi-normal modes.  

\ndt Additionally, we have to normalize the $\ket{HES}$ states since the ETH ansatz Eq.\eqref{eq:eth} that we compare with, is for normalized states (See Appendix \S\ref{app:norm}). We also replace ${| R_{ij} |}$ by its statistical average $1$, and then find $f(\omega)$ from the resulting answer by compensating for the entropic suppression. We demonstrate below in Fig. \ref{Fig4} the extracted $f(\omega)$ as a function of $\omega$. The average energy is fixed at $6.49$ in $\alpha'=1/2$ units. Our computing facilities allow for us to get upto ten different values of $\omega$. These correspond to $\left( N_1, N_2 \right) = \left( 22, 22 \right) ,  \left( 23, 21 \right), \left( 23, 22 \right),\left( 24, 20 \right) ,  \left( 24, 21 \right), \left( 25, 19 \right),\left( 25, 20 \right) ,  \left( 26, 18 \right), \left( 26, 19 \right),\left( 27,18 \right)$. Each of these elements involve averaging over a million amplitudes, which is further evaluated for different kinematical angles. For instance $N_1 = 25, N_2 = 20$ involves coarse-graining over 1227666 different matrix elements. Initially $f(\omega)$ stays at a constant for $\omega \rightarrow 0$ and then there is a range for which it decays exponentially with $\omega$ ( see left panel in Fig. \ref{Fig4} ). This is the expectation from ETH \cite{ETH-Chaos}. This expectation is met for angles close to $\chi = \pi/2$ which we saw dominates the diagonal ensemble. For other angles though the $e^{-S/2}$ suppression is still present, $f(\omega)$ does not show any monotonic decay ( see right panel in Fig. \ref{Fig4} ).

\begin{figure} 
\centering
\rotatebox{0}{\includegraphics*[width= 0.48 \linewidth]{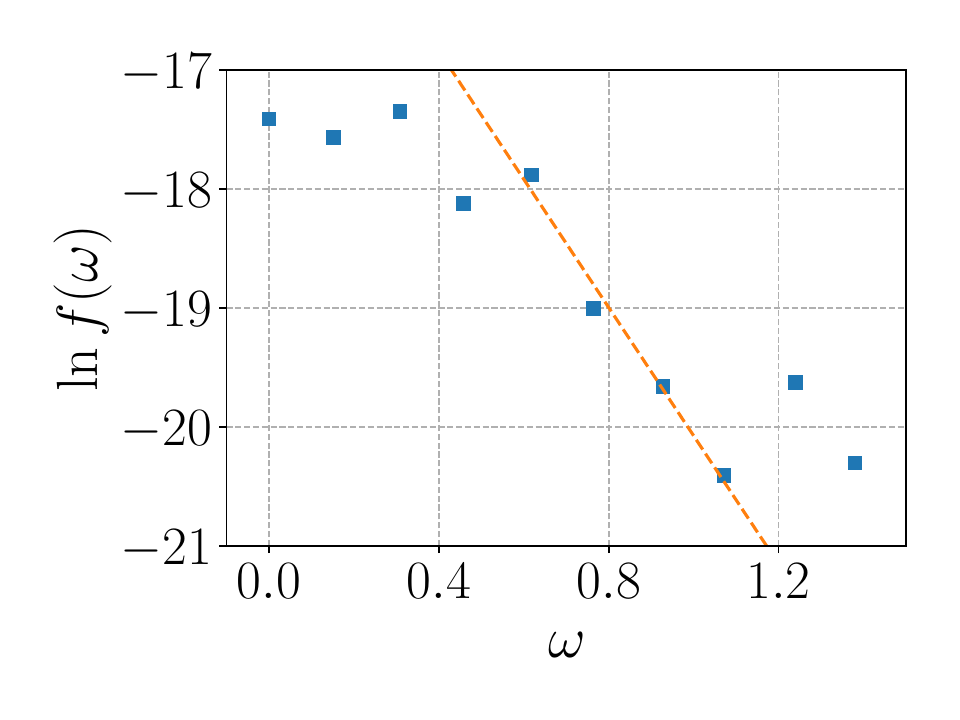}}
\rotatebox{0}{\includegraphics*[width= 0.48 \linewidth]{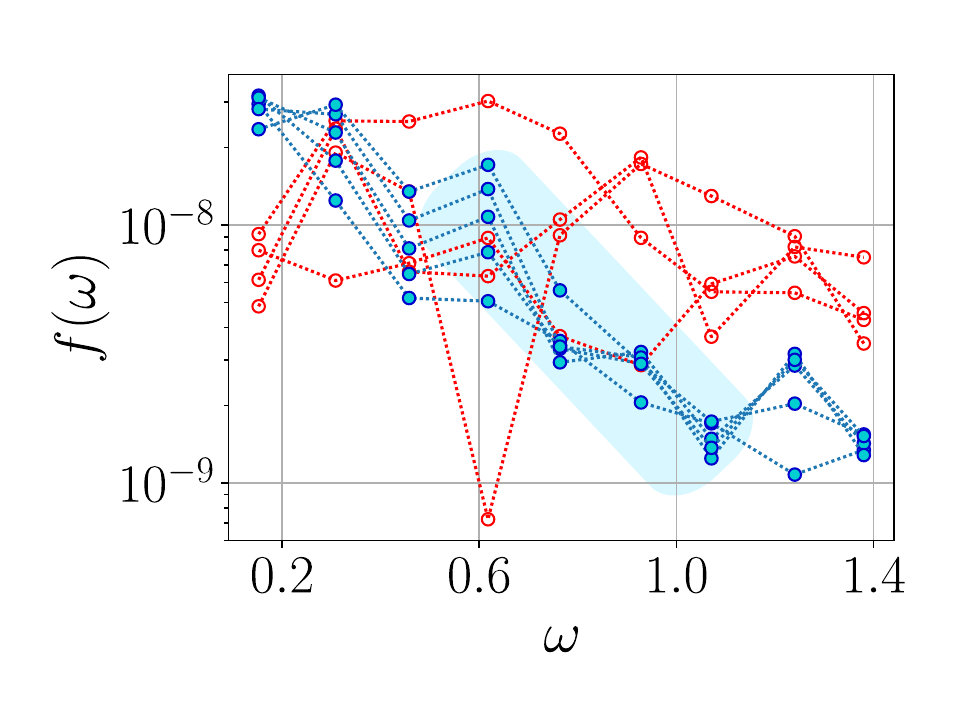}}
\caption{Left Panel : Plot of $\log f(\omega)$ as a function of $\omega$ for $\chi \approx \pi/2$. At intermediate $\omega$ we see an exponential decay (red dashed line). Right Panel :  Plot of $f(\omega)$ vs. $\omega$ for different values of $\chi$ using the ETH ansatz for the coarse-grained matrix element Eq.\eqref{eq:tilde} as a function of $\omega$. The blue banded data shows exponential decay for intermediate $\omega$, while the red data points show fluctuations. The blue points consist of $\chi = \left( 1.42, 1.47, 1.52, 1.57, 1.62 \right)$ radians all $ \approx \pi/2$, whereas the red set corresponds to $\chi = \left( 0.20, 0.70, 0.177, 2.50 \right)$ radians.}\label{Fig4}
\end{figure}

	\section{Computation of $HES_1(p_1,\zeta_1)+T_2(p_2)  \to HES_3(p_3,\zeta_3)+T_4(p_4)$}
	\label{sec:HHTT}
In this section we compute the amplitude corresponding to the process as depicted below in Fig.\ref{fig5a}. 
\begin{figure} 
\centering
\rotatebox{0}{\includegraphics*[width= 0.5 \linewidth]{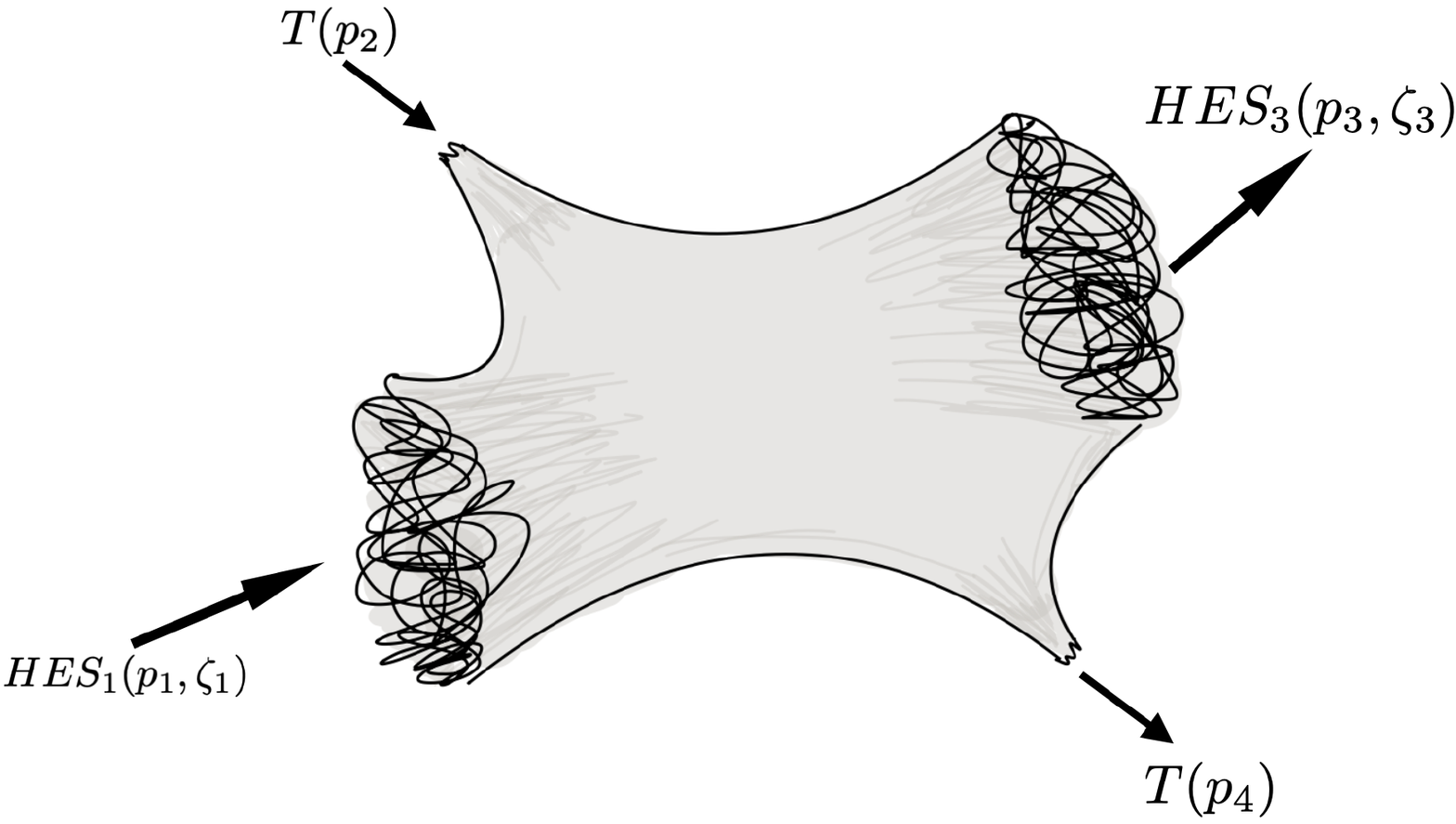}}
\rotatebox{0}{\includegraphics*[width= 0.42 \linewidth]{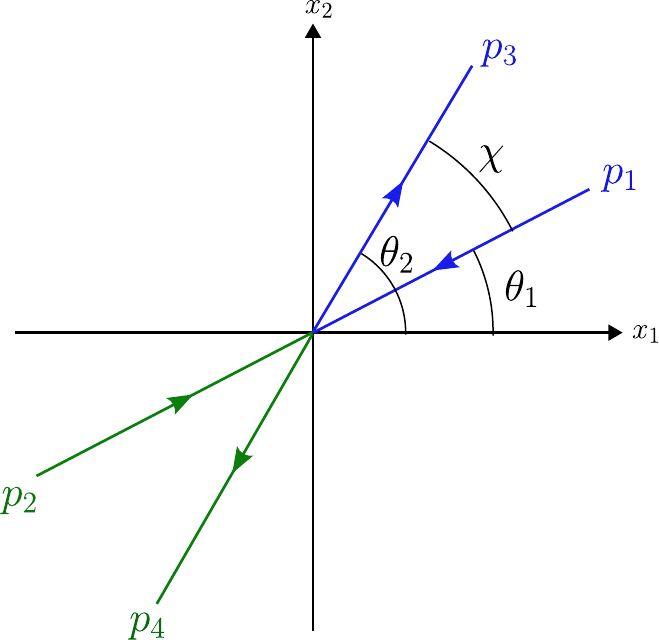}}
\caption{Left Panel : The $2\leftrightarrow 2$ scattering set-up considered. Right Panel : Kinematics indicating the scattering angles, green arrows indicate tachyons, while blue the $HES$.}\label{fig5a}
\end{figure}
\ndt Since this is a 4 point amplitude, it involves one integral over the cross-ratio. The integrand involves contractions between 4 vertex operators. 	
	\begin{align}
		e^{\cal L} =\bigg\langle :\prod_{r_1} &\left( \sum_{m_1}^{r_1} \frac{i }{(m_1 - 1)!} \zeta_1 \cdot \partial^{m_1} X_1 S^1_{r_1-m_1} \right) e^{i p_1 \cdot X_1}:: \prod_{r_3} \left( \sum_{m_3}^{r_3} \frac{i }{(m_3 - 1)!} \zeta_3 \cdot \partial^{m_3} X_3 S^3_{r_3-m_3} \right) \nonumber \\ &e^{i p_3 \cdot X_3}: : e^{i p_2 \cdot X_2 }: : e^{i p_4 \cdot X_4 }:\bigg\rangle
	\end{align}
Once again, we choose the constituent photons for $\ket{HES_1}$ and $\ket{HES_3}$ parallel to each other, i.e., $q_3 = -\gamma q_1$. We also naturally have the relations:
	\begin{align}
		 \zeta_1 \cdot \zeta_3 &= \lambda_1 \cdot \lambda_3, \,\,\
		\zeta_i \cdot q_j = 0, \,\,\, \lambda_i^2 = 0.
	\end{align}
Next we carry out the contractions. Then using $SL_2$ invariance we fix, $z_1 = 0, z_2= x$ and $z_3 =1$ and $z_4 = \infty$. The four point amplitude is obtained by integrating over $x$ from $-\infty$ and $\infty$. If we choose kinematics such that $q_i \cdot p_2 = 0$ as well as $\zeta_i \cdot p_2 = 0$, then we have:
	
	\begin{align} 
		{\cal A}_4 &= (-1)^{N_1 + N_3}  \int \, dx \,\,x^{N_1+N_3-2-p_1 \cdot p_3 - p_3 \cdot p_2}(1 -x )^{p_3 \cdot p_2 } \nonumber \\ &\sum_{k=0}^{\lbrace J_1,J_3 \rbrace_{min}} \sum_{\substack{\lbrace a(k)\rbrace \subset \lbrace r_{1} \rbrace\\ \lbrace b{(k)}\rbrace \subset \lbrace r_{3} \rbrace\\ \text{ordered}}} \frac{1}{k!} \prod_{i=1}^{k} \sum_{m_1,m_3=1}^{a_i,b_i}  \bigg[  \frac{(-1)^{m_3 +1 }(m_3 + m_1 -1)! \lambda_1 \cdot \lambda_3 }{(m_1 - 1)! (m_3 - 1)!  } \nonumber \\
		&\times S_{a_i-m_1} \bigg( \frac{a_i}{s_1} \left(  q_1 \cdot p_3 \right) \bigg)\times S_{b_i-m_3} \bigg( \frac{b_i}{s_3} \bigg(  (-1)^{s_3}q_3 \cdot p_1 \bigg) \bigg)\bigg] \nonumber\\
		&\times \prod_{l_1 \in \overline{\lbrace a(k) \rbrace}} \Bigg[ \sum_{n_1=1}^{l_1}\bigg(\zeta_1 \cdot p_3  \bigg)S_{l_1-n_1} \bigg( \frac{l_1}{s_1} \left(  q_1 \cdot p_3 \right) \bigg)\bigg] \nonumber \\
		&\times\prod_{l_3 \in \overline{\lbrace b(k) \rbrace}} \Bigg[ \sum_{n_3=1}^{l_3}\bigg((-1)^{n_3} \zeta_3\cdot p_1  \bigg) 
		S_{l_3-m_3} \bigg( \frac{l_3}{s_3} \bigg(  (-1)^{s_3}q_3 \cdot p_1\bigg) \bigg)\bigg]
	\end{align}

\begin{figure}[htbp]
\centering
\includegraphics[width=15cm]{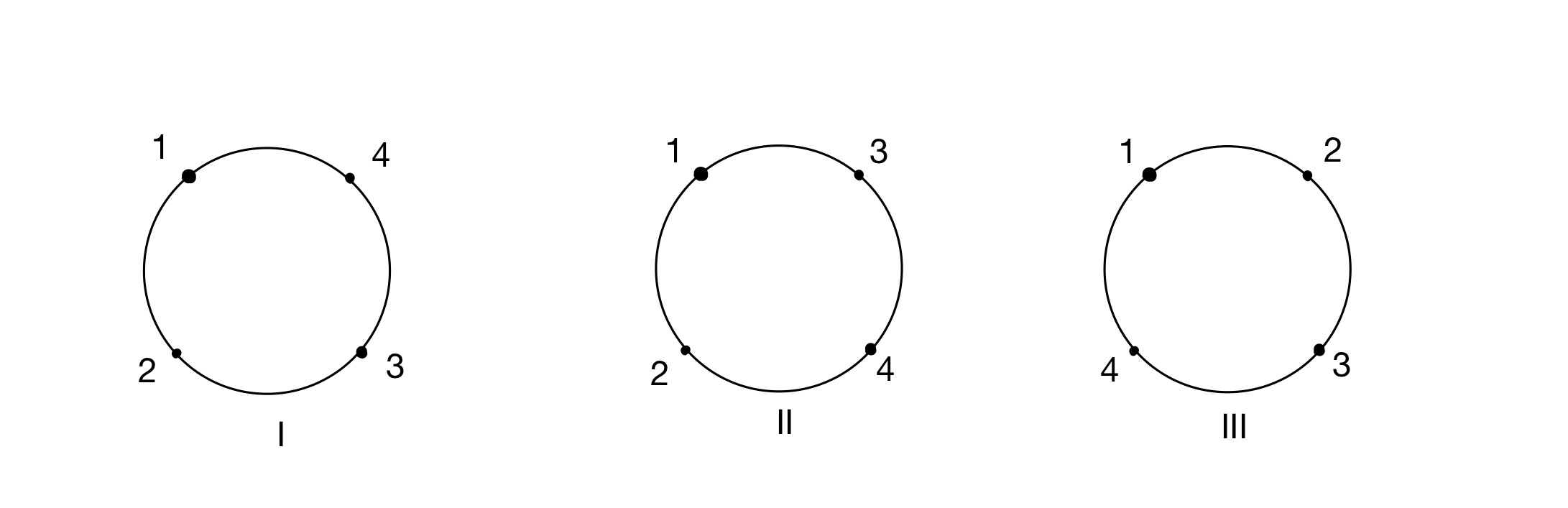} 
\caption{The process can happen in 3 different channels. Interchanging the positions of any two strings will correspond to a different channel. There can be six different diagrams, but $z_1=0$, $z_3=1$ and $z_4=\infty$ is fixed so the other $3$ channels are equivalent to these $3$.}
\label{process}
\end{figure}
\ndt The range of the above integration depends on the choice of channel. The integrals give rise to beta functions. The total amplitude therefore is a sum of different weighted beta functions, corresponding to different channels. In figure \ref{process} the string configurations for different channels are shown. The full amplitude takes the compact form: 
	\begin{equation} \label{hhtt-amp}
	\boxed{
		{\cal A}_4 =f_{N_{tot}}(s,t) \sum_{k=0}^{min \lbrace J_1, J_3 \rbrace} \frac{1}{k!} \sum_{\substack{\lbrace a(k) \rbrace \\ \lbrace b(k) \rbrace } } \left(\prod_{i=1}^{k} P(a_i,b_i) \prod_{l_1 \in \overline{\lbrace a(k) \rbrace}}Q_1(l_1) \prod_{l_3 \in \overline{\lbrace b(k) \rbrace}}Q_3(l_3) \right)}, 
	\end{equation}
where $P(a_i,b_i), Q_1$ and $Q_3$ are given in Appendix \S\ref{app:PQ}. The overall multiplicative factor $f_{N_{tot}}(s,t)$ is given by the following integral formula that we evaluate using the analytic continuation of beta function as:
\begin{align}
f_{N_{tot}}(s,t) &=(-1)^{N_{tot} }\Bigg[   \int\limits_{-\infty}^{0}  +  \int\limits_{0}^{1}  + \int\limits_{1}^{+\infty}  \Bigg] \,dx\,\, x^{N_{tot}-2-p_1 \cdot p_3 - p_3 \cdot p_2} (1-x)^{-p_3 \cdot p_2 } \nn \\
		&=(-1)^{N_{tot}} \Bigg[(-1)^{p_2 \cdot p_3}B(3-N_{tot} +p_1 \cdot p_3,p_2 \cdot p_3+1)+B(N_{tot}-1-p_1 \cdot p_3-p_2 \cdot p_3,p_2 \cdot p_3+1) \nn \\
		&+(-1)^{N_{tot}-1-p_1 \cdot p_3-p_2 \cdot p_3}B(N_{tot}-1-p_1 \cdot p_3-p_2 \cdot p_3,3-N_{tot}+p_1 \cdot p_3)\Bigg],
\end{align}
where $N_{tot}= N_1 + N_3$. Here, $s=-(p_1 + p_2)^2$ and $t=-(p_1 + p_3)^2$ are the Mandelstam invariants. One can see the structural similarity between Eq.\eqref{hhtt-amp} and the $HHT$ scattering amplitude, Eq.\eqref{HHT amp} ; thus for the given kinematical choices we find
 \begin{align}
 {\cal {A}}_4(p_i, \zeta_i )  \sim {\cal {A}}_3(p_i,\zeta_i)  f_{N_{tot}}(p_i),\label{eq:amp4amp3}
 \end{align}
i.e., the three point amplitude ${\cal{A}}_3$ acts like a dressing of a scalar amplitude to give the required 4 point amplitude with both momenta as well as the polarization vectors. 

\subsection{Evaluation of $HHTT$ amplitude}	
	Metric signature is once again $(-,+,+\dots +)$. The on-shell condition is $p_i^2 = - M_i^2$. Furthermore we take all the momenta to be ingoing. We choose the combined centre-of-mass frame of $H_1$ and $T_2$. These are ingoing states. The outgoing states are $H_3$ and $T_4$. The states $H_1$ and $H_3$ have rest masses $\sqrt{2(N_1-1)}$ and $\sqrt{2(N_3-1)}$ respectively. Maintaining the constraints we explicitly parametrize the kinematical variables in Appendix \S \ref{app:kin2}. The DDF photon momenta associated with $H_1$ and $H_3$ are related once again as: $q_3 = -\gamma \, q_1$. The amplitude can be expressed in terms of the Mandelstam $s = -\left(p_1 + p_2\right)^2, t = -(p_1 + p_3)^2$ or in terms of $s, \chi$. We note down the relationships between these variables in terms of the rest energy of the incoming probe $T_2$ (denoted by $e_0$) as: 
		\begin{align}
			s&=\left( \sqrt{e_0^2+2 N}+e_0 \right)^2, \,\,\, 
			t=4 \left(e_0^2+2\right) \cos ^2\left(\frac{\chi}{2}\right), \,\,\,
			\gamma=\frac{\sqrt{e_0^2+2 N}+e_0}{\sqrt{e_0^2+2 N}-e_0 \cos \chi}. 	\label{ghhtt}
	\end{align}
	
\ndt When $\zeta_1 \cdot \zeta_3=0$ and $N_1=N_3=N$, the amplitude can be expressed in terms of $s , \chi$:	
\begin{align}			
			\mathcal{A}_4 &=f_{2N} \left(s,t \right) \prod_{\lbrace r_1 \rbrace}\left(-\left(\sqrt{\frac{(s-2 N)^2}{8s}+1} \right) \sin (\chi)\prod_{j_1=1}^{r_1-1} \bigg(\frac{r_1/ \gamma -j_1}{j_1}\bigg) \right)^{n_{r_1}}\nn \\
			&\times \prod_{\lbrace r_3 \rbrace}\left(-\frac{s \left(\sqrt{\frac{(s-2 N)^2}{2s}+4} \right) \sin (\chi)}{ (2 N-s) \cos (\chi)+2 N+s} \prod_{j_3=1}^{r_3-1} \bigg(\frac{r_3\gamma -j_3}{j_3}\bigg) \right)^{n_{r_3}}
		 \label{HHTT amp}
	\end{align}

	\section{$HHTT$ in different regimes and its statistics}
\label{sec:	HHTT-average}
In this section we look at various asymptotic regimes where the above amplitude exhibits interesting behaviours. We also demonstrate its chaotic nature in the probe limit, as well as, point out how it realizes a notion of typicality.

\subsection{Fixed angle high energy scattering}
This is the limit when $ s \gg 2N$, and $\chi$ is fixed. The various factors become:  
\begin{align}
	f_{2N}(s,\chi) &\simeq B \left(-\frac{s}{2},\frac{s}{2}  \sin^2 \tfrac{\chi}{2} \right)-e^{-\frac{1}{2} i \pi  s}B \left(-\frac{s}{2},\frac{s}{2} \cos^2 \tfrac{\chi}{2}\right)+e^{\frac{1}{2} i s \pi   \sin^2 ({\small {\chi}/{2}})} B \left(\frac{s}{2}  \sin^2 \tfrac{\chi}{2} ,\frac{s}{2} \cos^2 \tfrac{\chi}{2} \right),  \label{eq:larges} \\
	\gamma &\sim \frac{1}{\sin ^2 (\frac{\chi}{2})}, \,\,
	\zeta_1 \cdot p_3   \sim -\frac{\sqrt{s}}{2 \sqrt{2}} \sin(\chi), \,\,\, \zeta_3 \cdot p_1 \sim -\sqrt{\frac{s}{2}} \cot(\frac{\chi}{2}).\nn
\end{align}
The $4$ point amplitude in this limit becomes : 
\begin{empheq}[box=\fbox]{align}
	\mathcal{A}_4^{\text{fixed }\chi } &= f_{2N}(s,\chi) \bigg( -  \frac{\sqrt{s}}{2 \sqrt{2}} \sin \chi \bigg)^{J_1}  \bigg(  \sqrt{\frac{s}{2} } \cot \tfrac{\chi}{2} \bigg)^{J_3} \nonumber \\
	&\times\prod_{\lbrace r_3 \rbrace}\left( \frac{\Gamma(r_3 \csc^2 \tfrac{\chi}{2})\Gamma(r_3 - r_3 \csc^2\tfrac{\chi}{2})}{\Gamma(r_3)}  \sin \left( \pi r_3 \csc^2 \tfrac{\chi}{2} \right) \right)^{n_{r_3}}  \label{A s large} \\
	&\times \prod_{\lbrace r_1 \rbrace}\left( (-1)^{r_1}\frac{\Gamma(r_1 - r_1\cos^2 \tfrac{\chi}{2})\Gamma(r_1 \cos^2\tfrac{\chi}{2})}{\Gamma(r_1)} \sin \left(\pi r_1 \cos^2 \tfrac{\chi}{2} \right)\right)^{n_{r_1}}.\nn
\end{empheq}
Note, that when $N \gg 1$, this limit also satisfies the condition $s > |t| \gg N $ for all values of the scattering angle $\chi$. 
In \cite{BF2} the $HTTT$ amplitude was also computed in this limit. In their set-up, two tachyons come in head-on and the resulting scattering angle is $\theta$ : 
\begin{align}
\mathcal{A'}_4^{\text{fixed } \theta } &=  f'_{N}(s,\theta) \bigg( - \sqrt{s} ( 1 - \tfrac{1}{2} \sin \theta ) \bigg)^J \prod_{\lbrace r \rbrace} \bigg( \frac{ \Gamma(r \sin \theta ) \Gamma( r -r\sin \theta ) }{\Gamma(r) } \sin \left(\pi r \sin \theta \right) \bigg)^{n_r}.
\end{align}
 The overall $s$ dependent factor in the forward scattering limit for $HTTT$ is $\left( \sqrt{s} \right)^{J}$ whereas, for $HHTT$ it is $\left( \sqrt{s} \right)^{J_1 + J_3}$. Furthermore the structure of the factors appearing within the product is also similar. The dressing factor doubles in the $HHTT$ case, corresponding to the two $\ket{HES}$ states. 

\subsection{Regge limit}

The Regge limit corresponds to $s \gg |t|$ irrespective of any bound on the parameter $N$. In Regge limit the various ingredients in the amplitude, Eq.\eqref{HHTT amp} behaves as:

\begin{align}
	\gamma &\simeq 1+ \mathcal{O} \left(\frac{|t|}{s}\right), \,\,
	f_{2N}(s,t) \simeq {\cal {A}}_{Ven}^{Regge} =\Gamma(-\tfrac{t}{2}-1) s^{\tfrac{t}{2} +1}, \,\,
	\zeta_1 \cdot p_3 \simeq \zeta_3 \cdot p_1 \simeq \sqrt{ \frac{|t|}{2}},  \\
	\mathcal{A}_4^{Regge} & \simeq {\cal {A}}_{Ven}^{Regge} \prod_{r_1}((-1)^{r_1}\zeta_1 \cdot p_3)^{n_{r_1}} \prod_{r_3}(\zeta_3 \cdot p_1)^{n_{r_3}} = (-1)^N\Gamma(-\tfrac{t}{2}-1) s^{\tfrac{t}{2} +1 } \left( \frac{|t|}{2}\right)^{\frac{J_1+J_3}{2}}.\nn
\end{align}
The exponent is consistent with the expected linear Regge trajectory. 

\subsection{Probe limit : $\alpha' s \sim  N \gg 1$}
\label{subsec:probe}
In case of classical scattering processes, the fractal structure in the scattering data is considered to be a marker of chaos. For example, one may consider a low-energy particle being scattered by some potential and plot the angle $\theta$ of the outgoing particle (i.e, the output parameter) as a function of the incoming impact parameter $b$. If the scattering is a chaotic one, then $\theta(b)$ will have regions in $b$ where the data will be speckled (i.e, regions of dense fluctuations). As one zooms into these dense regions, again self similar dense scattering data appears. Existence of this kind of fractal structure in the scattering data are considered as the marker of transient chaos \cite{seoane2012new}. In \cite{Hashi} such a set-up was investigated in the context of $HHTT$ scattering where the outgoing angle for a fixed ingoing one: $\theta'(\theta)$ was selected from the largest residue in the amplitude at fixed $s$. The strategy was implemented for fixed partitionings of $\ket{HES}$ states, hence there was no coarse-graining involved. Our strategy however is to implement coarse-graining, in order to make the notions of chaos and thermalizations emergent in the many-body context. 

\begin{figure} 
	\centering
	\includegraphics[scale=0.3]{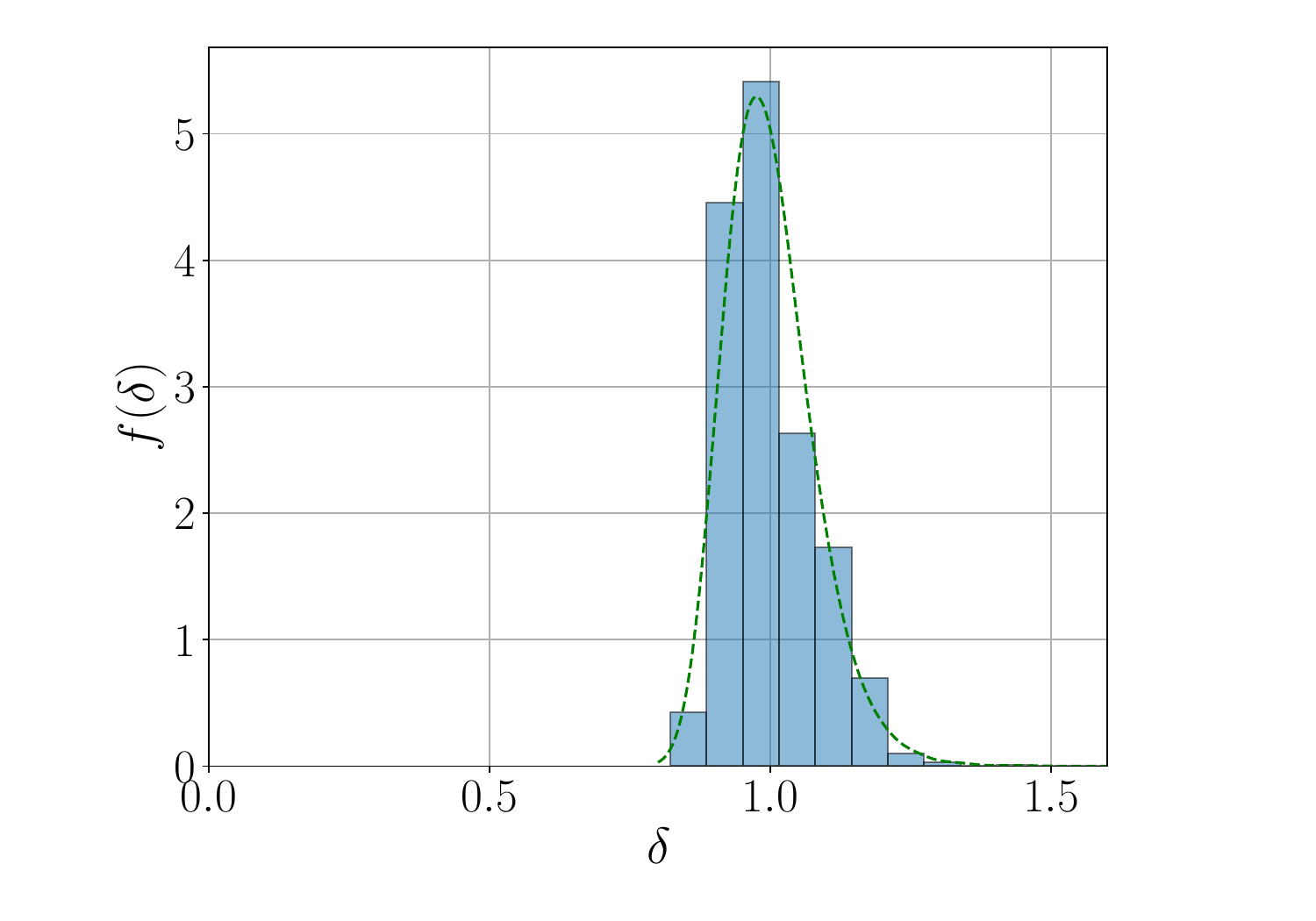}
	\includegraphics[scale=0.3]{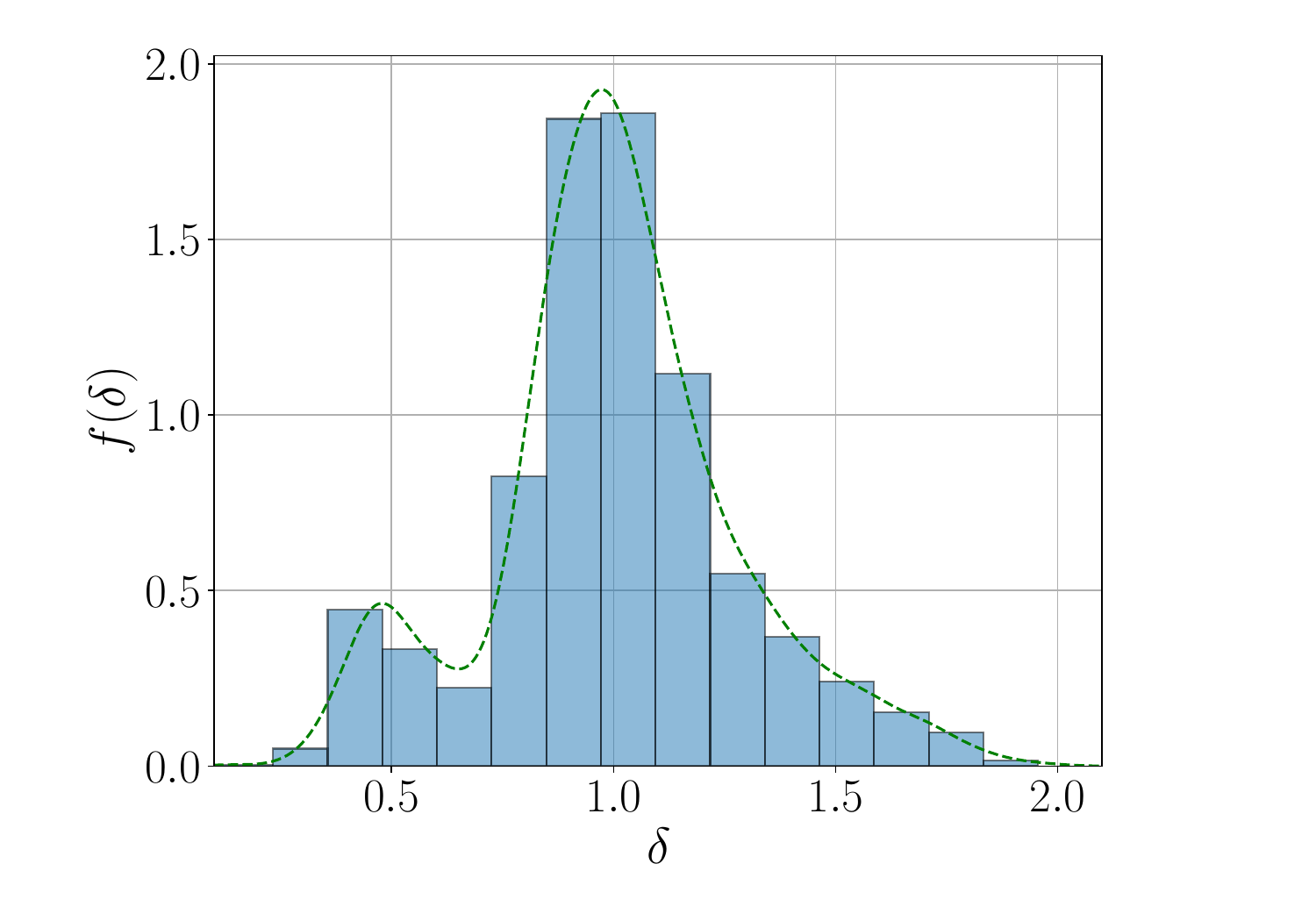}
	\caption{Nearest level spacing distribution of the residue  $\mathcal{A}_4^{\text{res}}$ at probe limit for two cases. Left Panel: $\zeta_1 \cdot \zeta_3 = 0$ and $N=20$, Right Panel: $\zeta_1 \cdot \zeta_3 \neq 0$ and $N=10$. Both of these show level repulsion. }
	\label{figHHTTprobeRes}
\end{figure}

\ndt The classical analysis demands that we consider scattering a light particle by some potential. In terms of the 4-point $HHTT$ scattering, this can be interpreted as a low-energy string scattering against the $\ket{HES}$ at much higher energies. When $s \, \alpha' $ is of order $N$ which is taken to be large, this corresponds to  $N \gg \alpha' e^2$. Here $t$ is fixed via the fixed scattering angle $\chi$, while we take the large $s$ limit. This means that the tachyon probe comes in with very low energy ($p_2^0 = e_0$) compared to the $\ket{HES}$ target. For $\alpha' = \tfrac{1}{2}$ and  $N_1 = N_3 = N$ the Veneziano amplitude portion behaves as,
\begin{align}
	f_{2N}(s,t)&=  \int\limits_{-\infty}^{+\infty}  x^{2N-2-p_1 \cdot p_3 - p_3 \cdot p_2}(1 -x )^{p_3 \cdot p_2}
	\sim  -\frac{4(1 + 2 \cos \chi)}{s-2N}  + {\cal{O}}\left((s-2N)^0 \right).
\end{align}
The $HHTT$ amplitude has a simple pole at $s \to N/\alpha'$, whose residue is:
\begin{itemize}
	\item For $\zeta_1 \cdot \zeta_3 = 0$,
	\begin{align}
		\mathcal{A}_4^{\text{res}} |_{s \to 2N}  \sim  (1 + 2 \cos \chi) (- \sin \chi)^{J_1+J_3}  (-1)^N ,  \label{eq:A4probe}
	\end{align}
	\item For $\zeta_1 \cdot \zeta_3 \neq 0$,
	\begin{align}
		\mathcal{A}_4^{\text{res}} |_{s \to 2N}  \sim  (1 + 2 \cos \chi) (- \sin \chi)^{J_1+J_3}  (-1)^N \sum_{k=0}^{\text{min} \lbrace J_1, J_3 \rbrace} \frac{( \sin \chi)^{-2 k}}{k!} \sum_{\substack{\lbrace a(k) \rbrace \\ \lbrace b(k) \rbrace } } \prod_{i=1}^{k} \delta _{a_i,b_i}  (-a_i) . 
	\end{align}
\end{itemize}
It is to be emphasized that even though the peak statistics of both these residues show level repulsion  (see Fig. \ref{figHHTTprobeRes}), they are far away from the Wigner surmise, and hence from the universal RMT predictions. We also observe that similar level statistics have been observed in the probe scalar spectrum in toy models of fuzzballs \cite{suman}. 

 Also, we do not find any randomness with the position of the largest residue within the $\ket{HES}$ ensemble. In this sense our results are consistent with absence of transient chaos if we focus only on the largest residue as in \cite{Hashi}.

\ndt The $\zeta_1 \cdot \zeta_3 = 0$ answer in Eq.\eqref{eq:A4probe}  is very similar to the three point amplitudes $HHT$  we got for large $N$. So the full average of the amplitude is given in the same way as equation \eqref{hhtaverage} : 
\begin{align}
	\langle \mathcal{A}_4\rangle_F  &\simeq -\frac{4(1 + 2 \cos \chi)}{s-2N}  \exp \Bigg[4\sqrt{N}\bigg( \sqrt{\text{Li}_{2}(-\sin \chi)}-\frac{\pi}{\sqrt{6}}\bigg) \Bigg].
	\label{hhttave}
\end{align}
Similar to the discussion of \S\ref{sec:HHT-average} we can also consider the diagonal part: 
\begin{align}
	\langle \mathcal{A}_4\rangle_D  &\simeq -\frac{4(1 + 2 \cos \chi)}{s-2N} \exp \Bigg[2\sqrt{N}\bigg( \sqrt{\text{Li}_{2}(\sin^2 \chi)}-\frac{\pi}{\sqrt{6}}\bigg) \Bigg].
	\label{hhttave}
\end{align}
Therefore the probe limit is also sensitive to the thermal scales of the problem just as the three point function is.

\subsection{Numerically probing typicality}
\label{subsec:HHTT-numerics}
	In this subsection we study the different microstate-channel contributions in the $HHTT$ S-matrix amplitudes, Eq.\eqref{HHTT amp}. We consider the case $N_1 = N = N_3$, and investigate numerically the partition dependences. We define the relevant part of the amplitude as:
	\begin{align}  \label{A tilde}
		\tmcA_4 (\mcP_{i}, \mcP_{j}) = & \prod_{ r_1 \in \mcP_{i} }\left(-\left(\sqrt{\frac{(s-2 N)^2}{8s}+1} \right) \sin (\chi)\prod_{j_1=1}^{r_1-1} \bigg(\frac{r_1/ \gamma -j_1}{j_1}\bigg) \right) \nn \\
		&\times \prod_{ r_3 \in \mcP_{j}}\left(-\frac{s \left(\sqrt{\frac{(s-2 N)^2}{2s}+4} \right) \sin (\chi)}{ (2 N-s) \cos (\chi)+2 N+s} \prod_{j_3=1}^{r_3-1} \bigg(\frac{r_3\gamma -j_3}{j_3}\bigg) \right)
	\end{align}
	where $\mcP_{i}$ and $\mcP_{j}$ are partitions of $N_1$ and $N_3$ respectively; the partitions can contain repeated elements (hence the absence of $n_{r_1}$ and $n_{r_3}$ factors in the expression).

\ndt Here we study $\tilde{\mcA}_4 ( \mcP_{i}, \mcP_{j})$ for $N=5$ case for a range $s \gtrsim 2N $ to $s \gg 2N$.  Here the partitions are denoted as $
		\mcP_1 \equiv \{5\}, ~ 	\mcP_2 \equiv\{4,1\}, ~  \mcP_3 \equiv \{3,2\},
		~ \mcP_4 \equiv\{3,1,1\}, 
		~ \mcP_5 \equiv\{2,2,1\}, ~ \mcP_6 \equiv\{2,1,1,1\}, ~ \mcP_7 \equiv\{1,1,1,1,1\} $.

\ndt Below in Fig.\ref{Fig6} and in Fig.\ref{fig:N5 full SPlot} the S-matrix absolute amplitude is plotted for increasing values of $s$ starting from $s \gtrsim 2N$ for $\zeta_1 \cdot \zeta_3 = 0$ and $\zeta_1 \cdot \zeta_3 \neq 0$ respectively. At $s \simeq 2N$, $\tmcA_4 (\mcP_1, \mcP_1)$ has the highest amplitude. For $\zeta_1\cdot \zeta_3 =0$ case, with increasing $s$ the amplitude matrix profile shifts smoothly and for $s \gtrsim 2N$ the maxima saturates at $\tmcA_4 (\mcP_1, \mcP_7)$. For even greater values of $s$ the maximum amplitude shifts smoothly to $\tmcA_4 (\mcP_7, \mcP_7)$ for $s \gg 2N$. This feature continues to hold independent of $\chi$ as well as $N$. 
		\begin{figure} 
		\centering
		\includegraphics[scale=0.8]{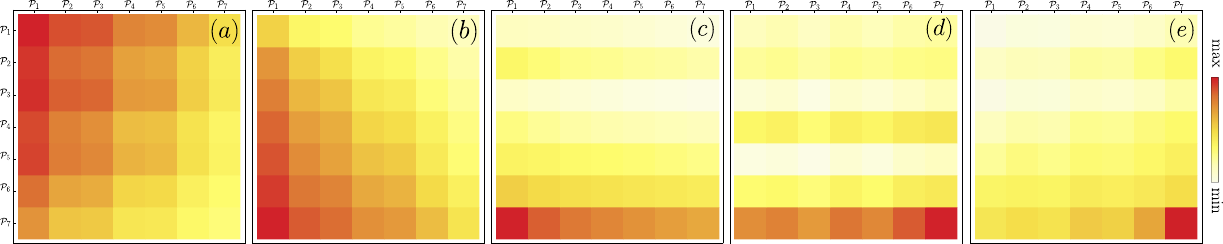}
		\caption{$|\tmcA_4 (N = 5,s,\chi = 1.2)|$ for the case $\zeta_1 \cdot \zeta_3 = 0$, plotted for the $s$ values (a) 10.137, (b) 11.1507, (c) 21.2878, (d) 44.6031, (e) 204.769. Column and row partitions correspond to partitions of $N_1$ and $N_2$ respectively. With increasing $s$ the maximum amplitude moves smoothly from $\tmcA_4 (\mcP_1, \mcP_1)$ in (a) to $\tmcA_4 (\mcP_1,\mcP_7)$ in (c). For larger $s$ the maxima smoothly saturates to $\tmcA_4 (\mcP_7, \mcP_7)$. }
		\label{Fig6}
	\end{figure}

\ndt The dominant microstate channels can be explained from Eq.\eqref{A tilde} in various regimes. At $s \simeq 2N$, one can expand $\tmcA_4$ to obtain (see Eq.\eqref{eq:A4probe}):
	\begin{align}  \label{A tilde s equals 2N}
		&\tmcA_4 ( \mcP_{i}, \mcP_{j}) \sim  \left(- \sin (\chi) \right)^{J_1 +J_3} (-1)^{N}
	\end{align}
	where, $ N = \sum_{r} r n_r, ~~~   J = \sum_{r} n_r$. Now, since we have $0 \leq \sin (\chi) \leq 1 $, $|\tmcA_4|$ is maximum when $(J_1 + J_3)$ is minimum. As $\mcP_1$ has $J = \min J = 1$, thus at $s \simeq 2N$ the $\tmcA_4 (\mcP_1, \mcP_1)$ component dominates. \\
	
\ndt For $s \gtrsim 2N$, we expand Eq.\eqref{A tilde} around $s = 2N (1 + \epsilon)$ with $\epsilon >0$ and keep till linear order in $\epsilon$. In this limit upto linear order in $\epsilon$ we obtain:
	\begin{align}
		\tmcA_4  (\mcP_{i}, \mcP_{j}) &\sim \prod_{\substack{ r_1 \in \mcP_{i} \\ r_1 \text{ distinct}}} \left(  -\sin(\chi) \left( 1 - \frac{\epsilon}{2} (1 + \cos (\chi)) ( \gamma_E + \psi(r_1) )   \right)  \right)^{n_{r_1}} \nn \\ &\prod_{\substack{ r_2 \in \mcP_{j} \\ r_2 \text{ distinct}}} \left(  -\sin(\chi) \left( 1 + \frac{\epsilon}{2} (1 + \cos (\chi)) (1 + \gamma_E + \psi(r_2) )   \right)  \right)^{n_{r_2}},
	\end{align}
where, $\gamma_E$ denotes the Euler number, $\psi(r) = \Gamma'(r)/\Gamma(r)$ is the digamma function. The magnitude of the first product is always less than $1$ whereas for the second it is larger than $1$. Thus for typical values of $\chi$, $\tmcA_4(s \gtrsim 2N)$ maximizes when $J_1 = \min J_1$ and $J_2 = \max J_2$; so the asymmetric corner channel $\tmcA_4(\mcP_1, \mcP_7)$ dominates the scattering process. 	\\ 

\ndt Next, for $s \gg 2N$, from Eq.\eqref{A s large} we have for non-zero values of $\chi$: $\tmcA_4 (\mcP_{i}, \mcP_{j}) \sim  \left( \sqrt{s} \right)^{J_1 + J_2}$. So the component with $J_1 = \max J_1$ and $J_2 = \max J_2$ dominates. Thus in case of $N = 5$, for $s \gg 2N$ the $\tmcA_4(\mcP_7, \mcP_7)$ has the maximum amplitude. \\

\ndt Lastly, for the $\chi = 0$, the $\chi$ dependent part in Eq.\eqref{A s large} seems to be diverging because of the $\cot (\chi/2)$ factor. But one can take the $\chi \to 0$ limit carefully: $\gamma \xrightarrow{\chi \to 0, ~ s \gg 2N} \frac{s}{2N} ; \,\, ~ \zeta_1.p_3 \xrightarrow{\chi \to 0, ~ s \gg 2N} \frac{\chi  \sqrt{s}}{2 \sqrt{2}}; \,\, ~ \zeta_3.p_1 \xrightarrow{\chi \to 0, ~ s \gg 2N} \frac{\chi  s^{3/2}}{4 \sqrt{2} N}$. Therefore in the limit $s \gg 2N, \chi \to 0$ : 	
	\begin{align}
		 \tmcA_4 ( \mcP_{i}, \mcP_{j} )  &\sim  \prod_{\substack{ r_1 \in \mcP_{i} \\ r_1 \text{ distinct}}} \left(  \frac{\chi  \sqrt{s}}{2 \sqrt{2}}  \right)^{n_{r_1}}  \prod_{\substack{ r_2 \in \mcP_{j} \\ r_2 \text{ distinct}}} \left(  \frac{\chi  s^{3/2}}{4 \sqrt{2} N}  \frac{1}{\Gamma (r_2)} \right)^{n_{r_2}}  \sim s^{J_1/2} s^{3J_2/2}
	\end{align}	
	Again this is maximum for $J_1 = \max J_1$ and $J_2 = \max J_2$; thus the $\tmcA_4(\mcP_7, \mcP_7)$ dominates for all values of $\chi$ when $s \gg 2N$. 
	
	\begin{figure} 
		\centering
		\includegraphics[scale=0.8]{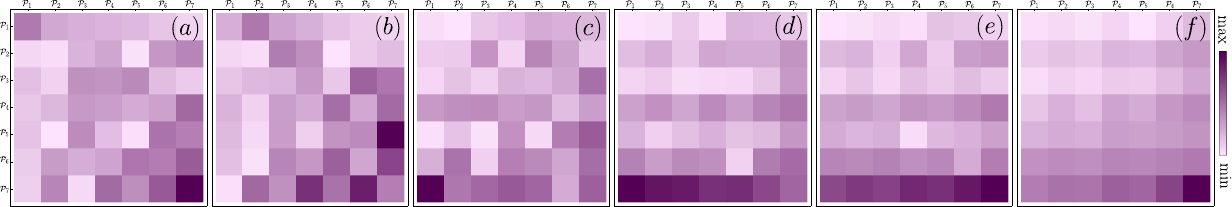}
		\caption{$|\tmcA_4 (N = 5,s,\chi = 1.2)|$ for the case $\zeta_1 \cdot \zeta_3 \neq 0$, plotted for the $s$ values (a) 10.137, (b) 15.2056, (c) 34.466, (d) 66.9048, (e) 93.2612, (f) 204.769. Column and row partitions correspond to partitions of $N_1$ and $N_2$ respectively. For $s \gtrsim 2N$, the amplitude is scattered throughout the partitions with no smooth transition. For larger $s$ the maxima smoothly saturates from $\tmcA_4 (\mcP_1,\mcP_7)$ in (d) to $\tmcA_4 (\mcP_7, \mcP_7)$ in (f). }
		\label{fig:N5 full SPlot}
	\end{figure}

\ndt Since the 4 point scattering amplitude has a dressing factor similar to the $HHT$ amplitude, Eq.\eqref{eq:amp4amp3}, the thermal characteristics of the three point amplitude get inherited into the four point function as well. It is then natural to consider the centre-of-mass energy scale $s$ to give rise to an effective temperature in this context. Our numerical observations indicate that at higher temperatures (large $s$) the microstate channel which dominates is the one made out of small integer partitions. We point out that this {\em typicality} is expected for the $\ket{HES}$ states which can be represented by integer partitionings or Young tableaux. Young tableaux follows the Bose-Einstein distribution \cite{vershik1996statistical}.  Excited descendants in 2D conformal field theories are also similarly labelled, and the same fact was used to establish typicality in \cite{Datta:2019jeo} by investigating stress-tensor correlators. Given, the Bose-Einstein distribution, the average occupation number $n_r$ of mode number $r$ at inverse temperature $\beta$, for a system of size $L$ is: 
\begin{align}
\vev{ n_r }_\beta &= \frac{1}{ e^{2\pi \beta/L r}  - 1 } \xrightarrow{\beta/L \ll 1 } \frac{L}{2\pi \beta r }. 
\end{align}
Hence at high temperatures the lowest mode number : $r = 1$ are dominantly occupied. This is consistent with the observed channel dominance at large $s$. Note, that an analysis of the dependence of the chaos in $ 2 \leftrightarrow 2$ scattering on the partitionings recently appeared in \cite{Savic:2024ock}.

\section{$HES$ form factor}
\label{sec:formfactor}

\begin{figure} 
\centering
\rotatebox{0}{\includegraphics*[width= 0.2 \linewidth]{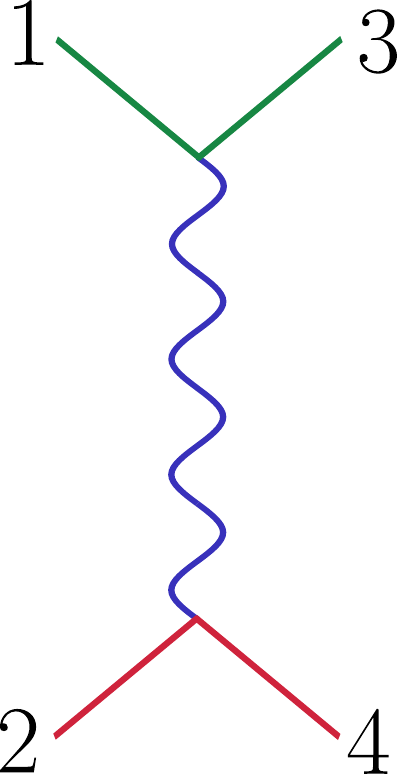}}\qquad \qquad \qquad \qquad \qquad 
\rotatebox{0}{\includegraphics*[width= 0.4 \linewidth]{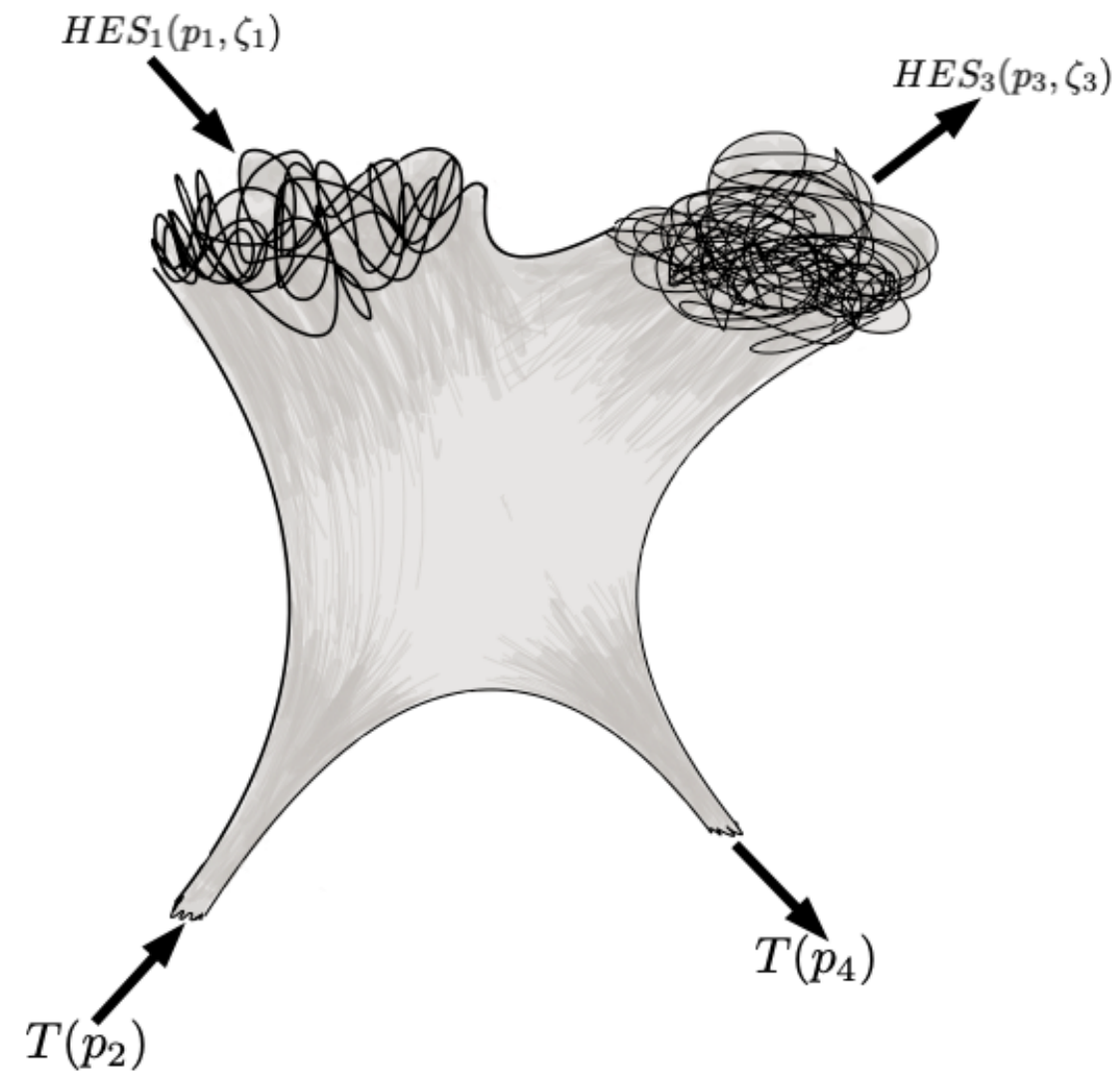}}
\caption{Left Panel : A QFT process with two different species of particles (show in different colours in the solid lines) and the internal propagator is shown with a wavy line.  Right Panel : The string interaction vertex at tree level.} \label{fig8}
\end{figure}

According to our kinematic choice the process we want to study is 
\begin{equation}
		HES_1(p_1)+T_2(p_2)  \to HES_3(p_3)+T_4(p_4).
\end{equation}
Consider a $2 \leftrightarrow 2$ scattering in a quantum field theory, with particles 1(2) identical to 3(4). Suppose that the form factor of particle 2 (hence 4) is known and we want to find out the form factor of particle 1 (hence 3). This information can be obtained by considering the scattering in the channel as depicted in Fig. \ref{fig8}. Similarly if the form factor of a tachyon is known, then the form factor of a HES state can be calculated from a similar vertex. In our case string $1$ and $3$ are HES and rest are tachyons. Hence we must consider the process in channel II in figure \ref{process}. A schematic diagram of the string interaction is showed in Fig. \ref{fig8}.
\ndt The momenta of the states are taken so that the $p_1  , p_2$ are ingoing states and $p_3  ,  p_4$ are outgoing states (therefore their energies are negative in accordance with our conventions). As we only consider the $t$-channel, the range of the integral in equation \eqref{hhtt-amp} would be $x \in (-\infty,0)$. In this section we set $N_1 = N_3 = N$. In the Regge limit the Veneziano factor of the equation \eqref{hhtt-amp} becomes,
\begin{align}
	f_{N_{tot}}(s,t) &\sim (-1)^{N+\frac{s}{2}+\frac{t}{2}} \left(\frac{s}{2}\right)^{1+\frac{t}{2}} \Gamma (-1-\frac{t}{2}).
	\end{align}
The fixed channel amplitude with $\zeta_1 \cdot \zeta_3=0$ is given by, 
\begin{align}
	{\cal{A}}_t &= (-1)^{N+\frac{s}{2}} \frac{ \Gamma (-1+N-\frac{s}{2}) \Gamma (-1-\frac{t}{2})}{\Gamma (-2+N-\frac{s}{2}-\frac{t}{2})} \prod_{\lbrace r_1\rbrace} \left(\frac{ \sqrt{-t} \sqrt{(s-2 N)^2+s t+8 s}}{\sqrt{2} \sqrt{(s-2 N)^2+8 s}} (-1)^{r_1} \prod_{j_1=1}^{r_1-1}\frac{r_1/\gamma -j_1}{j_1}\right)^{n_{r_1}}\nn\\
		\prod_{\lbrace r_3\rbrace}&\left(\frac{ \sqrt{-t} \sqrt{(s-2 N)^2+8 s} \sqrt{(s-2N)^2+s t+8 s}}{\sqrt{2} \left((s-2 N)^2-2 N t+s t+8 s\right)}  \prod_{j_3=1}^{r_3-1}\frac{r_3\gamma -j_3}{j_3}\right)^{n_{r_3}}.
\end{align}
We interpret Mandelstam variable $t$ as the transfer momentum at a $3$ point vertex, similar to QFT vertices i.e. $t=-Q^2$, where $Q=p_1+p_3$. In the Regge limit (when $s \gg |t| = Q^2$) we have:
\begin{align}
		\gamma&= 1+\frac{Q^2}{s}+O\left(s^{-2}\right), \,\, 
		\zeta_1 \cdot p_3 =\frac{Q}{\sqrt{2}}-\frac{Q^3}{\left(2 \sqrt{2}\right) s}+O\left(s^{-2}\right), \,\, 
		\zeta_3 \cdot p_1 &=\frac{Q}{\sqrt{2}}+\frac{Q^3}{\left(2 \sqrt{2}\right) s}+O\left(s^{-2}\right).
		\end{align}
Hence the amplitude simplifies to: 
\begin{align}
		\mathcal{A} &\sim (-1)^{\frac{s}{2}+\frac{t}{2}} \left(\frac{s}{2}\right)^{1+\frac{t}{2}} \Gamma (-1-\frac{t}{2}) \prod_{\lbrace r_1\rbrace} \left(\frac{Q}{\sqrt{2}}\right)^{n_{r_1}} \prod_{\lbrace r_3\rbrace} \left(\frac{Q}{\sqrt{2}}\right)^{n_{r_3}}
	\end{align}
At small momentum transfer $Q^2 \to 0$, there is a $Q^2=0$ pole in the amplitude with coefficient given as below,
\begin{align}
		\mathcal{A} &\sim \frac{1}{Q^2} (-1)^{\frac{s-Q^2}{2}}s \exp[-\frac{Q^2}{2} \log (\frac{s}{2})] \left(\frac{Q}{\sqrt{2}}\right)^{J_1+J_3} 
	\sim \frac{1}{Q^2}\mathcal{F}_c(Q^2) \mathcal{F}_{N,\lbrace n_{r_1}\rbrace,\lbrace n_{r_3}\rbrace}(Q^2).\label{eq106} 
	\end{align} 
In the above expression,  $$\mathcal{F}_c(Q^2)=s \exp[-\frac{Q^2}{2} \log (\frac{s}{2})]$$ is the tachyon vertex form factor and $\mathcal{F}_{N,\lbrace n_{r_1}\rbrace,\lbrace n_{r_3}\rbrace}(Q^2)=(-1)^{\frac{s-Q^2}{2}}\left(\frac{Q}{\sqrt{2}}\right)^{J_1+J_3}$ is the form factor for the $HES-HES$-massless intermediate particle vertex form factor. This is the leading order behavior of the quantity which depends on the partition of the $\ket{HES}$ state we are choosing. The average form factor can be obtained by \textit{summing} over the final microstates at level $N_3$ and by {\it averaging} over the initial microstates at level $N_1$ \cite{Manes:2003mw} : 

\begin{align}
		\langle |\mathcal{A}|^2 \rangle &= \frac{1}{\Omega(N_1)} \sum_{\{n_{r_1}\},\{n_{r_3}\}} {|\mathcal{A}|^2} =   \frac{1}{Q^4}|\mathcal{F}_c(Q^2)|^2 |\mathcal{F}_N(Q^2)|^2
\end{align}
We consider elastic limit with $N_1 = N_3 = N$. For all possible in states and out states, the average $4$ point amplitude becomes: 
\begin{align}
\vev{|{\cal A}|^2} &= \frac{1}{\Omega(N) } \sum_{\substack{\lbrace n_{r_1} \rbrace \\
				\lbrace n_{r_3} \rbrace}} \oint d\beta_1 d\beta_3 \,\, e^{ \beta_1 N +\beta_3 N }e^{- \beta_1 \sum_{r_1} r_1 n_{r_1} }e^{- 		                      \beta_3 \sum_{r_3} r_3 n_{r_3} } |\mathcal{A}_{(\lbrace r_1 \rbrace,\lbrace r_3 \rbrace,\zeta_1 \cdot \zeta_3=0 )}|^2
\end{align}
In the Regge limit of small momenta transfer we find using Eq.\eqref{eq106} : 
\begin{equation}
	\begin{split}
		\vev{|{\cal A}|^2} &=   \frac{ s^2 \exp[Q^2 \log (\frac{s}{2}) ]}{\Omega(N)\, Q^4}\sum_{\substack{n_{r_1}=0\\
				n_{r_3}=0}}^{\infty}\oint d\beta_1 d\beta_3 \,\, e^{ \beta_i N } \prod_{r_1=1}^{\infty} e^{- \beta_1 r_1 n_{r_1} }
		\left(\frac{Q^2}{2}	\right)^{ n_{r_1}}
		\prod_{r_3=1}^{\infty}e^{- \beta_3 r_3 n_{r_3} } \left(\frac{Q^2}{2}	\right)^{ n_{r_3}}\\
		&=   \frac{ s^2 \exp[Q^2 \log (\frac{s}{2}) ]}{\Omega(N) \, Q^4}  \oint d\beta_1 d\beta_3 \,\, e^{ \beta_i N } \prod\limits_{r_1 \in \{ N\} } 
		\frac{1}{ 1 -\frac{Q^2}{2} e^{-\beta_1 r_1}} \prod\limits_{r_3 \in \{N \} }\frac{1}{ 1 - \frac{Q^2}{2}e^{-\beta_3 r_3}} .
	\end{split}
\end{equation}
We proceed to evaluate the $\beta_1, \beta_3$ integrals using saddle point : 
%
\begin{equation}
	\begin{split}
		\vev{|{\cal A}|^2} &=  \frac{ s^2 \exp[Q^2 \log (\frac{s}{2}) ]}{Q^4} \exp \left[4\sqrt{N}\bigg\lbrace \sqrt{\text{Li}_{2}\frac{Q^2}{2})}-\frac{\pi}{2 \sqrt{6}}
		\bigg\rbrace \right]
	\end{split}
\end{equation}
Therefore we identify the effective excited state form factor to be : 
\begin{align}
	\mathcal{F}_N (Q^2)  &=  \exp \bigg[4\sqrt{N} \bigg( \sqrt{\text{Li}_{2} \left(\frac{Q^2}{2} \right)}-\frac{\pi}{2\sqrt{6}}\bigg)\bigg]. \label{eq111}
\end{align}
%
%
%
\subsection{Size of the $HES$ state } 
\label{subsec:size}
From the form factor we can obtain the spatial distribution of the HES target in $D-1$ dimensions using Fourier transform:
\begin{align} \label{form-furier}
	\rho (\mathbf{r}) &= \left(\frac{1}{2 \pi }\right)^{D-1}\int \text{d}^{D-1} Q \,\, \mathcal{F}_N (\mathbf{Q}^2) \, e^{i \mathbf{Q} \cdot \mathbf{r}} \nonumber \\
	&= \left(\frac{1}{2 \pi }\right)^{D-1} K_{D-3} \int_{0}^{\pi} \sin^{D-3} \theta d \theta \int_{0}^{\infty} d q~ q^{D-2}e^{i q r \cos \theta} \mathcal{F}_N (q^2)
\end{align}
We have introduced the $D-1$ dimensional polar coordinates. $K_{D-3}$ is the constant obtained by integrating the $D-3$ angular coordinates.  Note that from Eq.\eqref{eq111}, at $q=0$, since the dilogarithm function goes to zero, ${\cal F}_N$ becomes a constant. Therefore at zero momentum transfer the HES target appears like a point particle as $\rho(r) \sim \delta(r)$. To evaluate $\rho(r)$ for non-zero $q$ we further specialize to $D=4$  where $K_1=\int_{0}^{2 \pi} d \phi= 2 \pi$. 
\begin{align}
\rho(r) &= \frac{1}{4\pi^2} \int_0^\pi d\theta\,\, \sin \theta e^{i \, q r \cos \theta} \int_0^\infty dq\,\,q^2\exp \bigg[4\sqrt{N} \bigg( \sqrt{\text{Li}_{2} \left(\frac{q^2}{2} \right)}-\frac{\pi}{2\sqrt{6}}\bigg)\bigg] \nn \\
&= \frac{1}{2\pi^2 r} \text{Im} \bigg( \int_0^\infty dq\,\,q \exp \bigg[i \,q r + 4\sqrt{N} \bigg( \sqrt{\text{Li}_{2} \left(\frac{q^2}{2} \right)}-\frac{\pi}{2\sqrt{6}}\bigg)\bigg] \bigg) 
\end{align}
Note, that consistency requires that we focus on large $r$ regimes, i.e., $r \gg q^{-1}$, wherein we can use the saddle point approximation to evaluate the integral. We carry this out numerically and obtain profiles of $\rho(r)$ as shown in Fig.\ref{figrho} below. 
\begin{figure} 
\rotatebox{0}{\includegraphics*[width= 0.42 \linewidth]{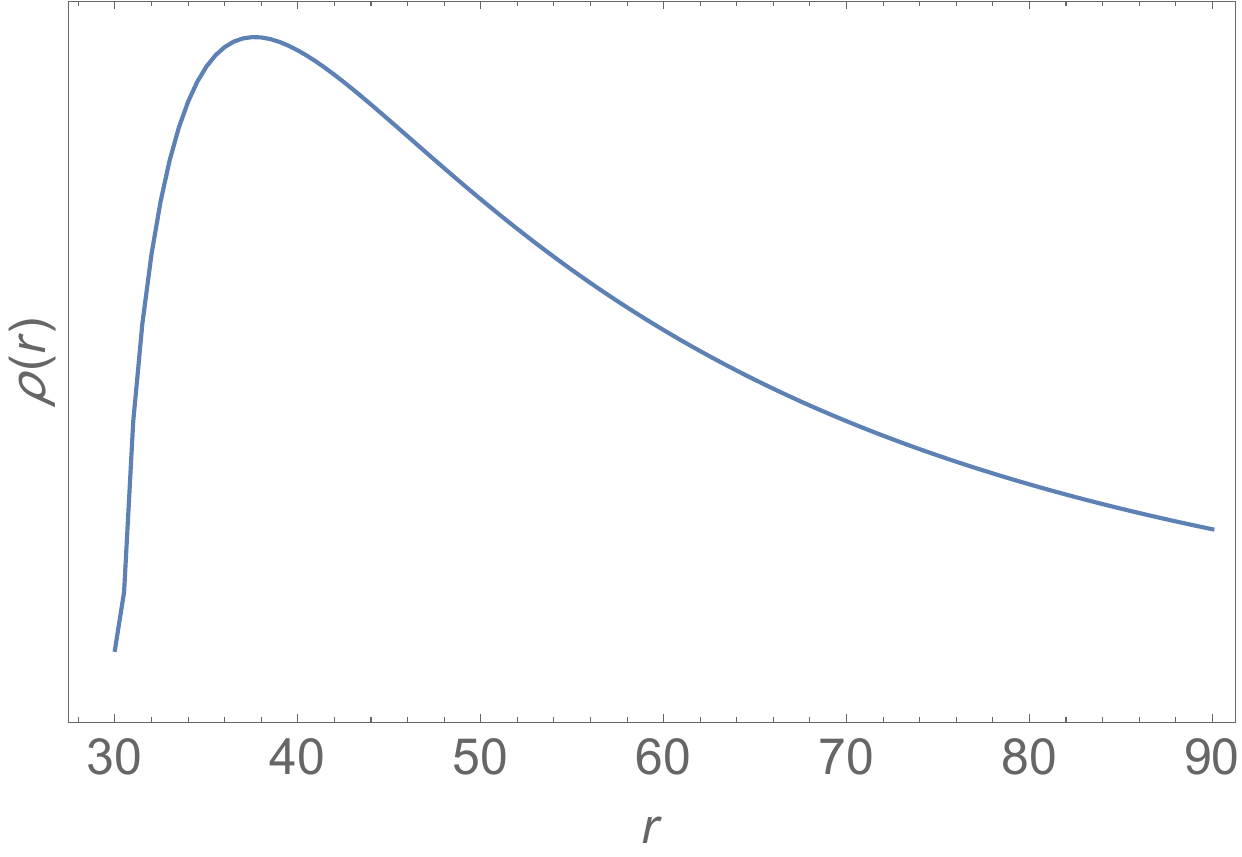}}
\rotatebox{0}{\includegraphics*[width= 0.46 \linewidth]{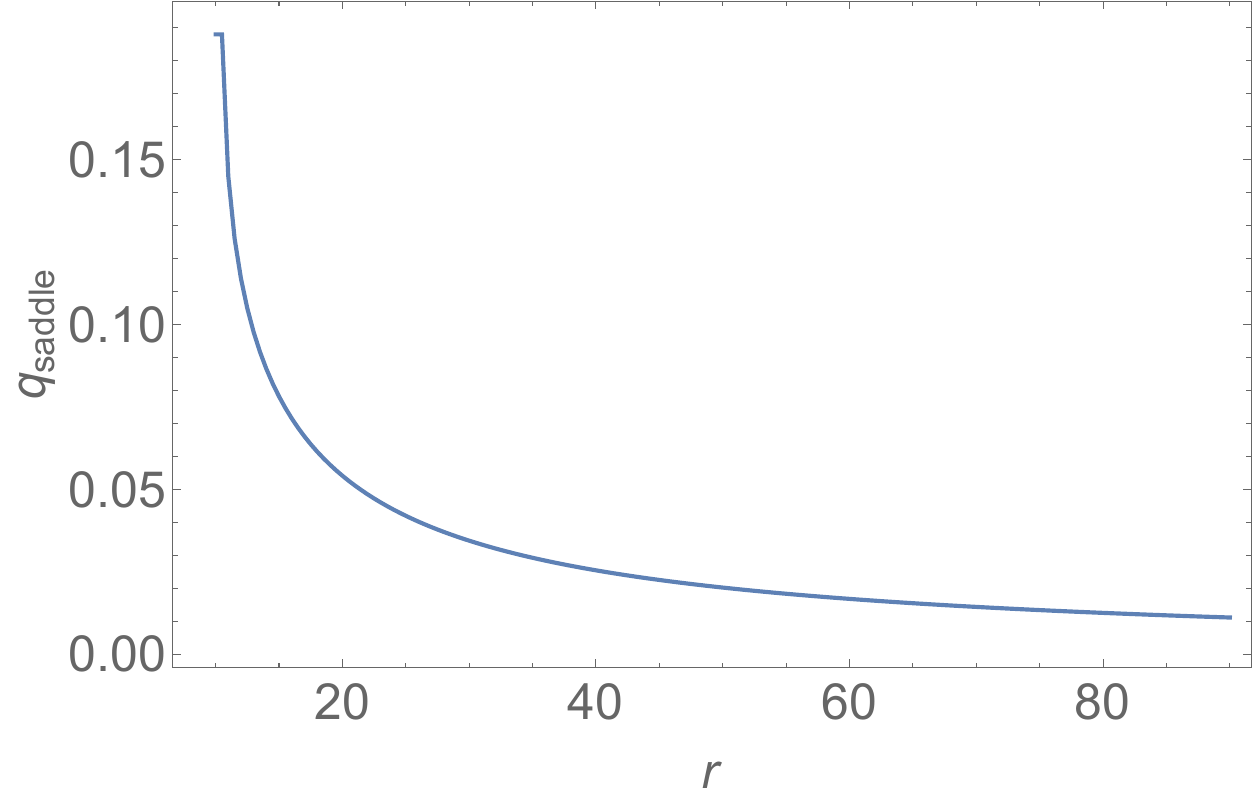}}
\caption{Left Panel : Profile of $\rho(r)$ as a function of $r$ for $N=50$. Right Panel : The value of the saddle point as a function of $r$.} \label{figrho}
\end{figure}
\ndt In the numerical evaluation we keep track of the consistency of the saddle point by tracking the value of $q_{\texttt{saddle}}$ as a function of $r$. The large $r$ region where the saddle point approximation has less error, we find that the target profile decays. For the Amati-Ruso set-up this form factor was computed in \cite{Manes:2003mw} and the distribution $\rho(r)$ was shown to exactly match with that of a random walk distribution \cite{Manes:2004nd}. In particular one obtains that $\vev{r^2}\sim L$ where $L$ is the total length of the string \cite{Mitchell:1987th}. The total string length $L$ being proportional to its mass goes like $\sqrt{N }\ell_s$. The intuitive picture is that the time-snap of the string is that of a random walk each of $\ell_s$ units, repeated $\sqrt{N}$ number of times. Therefore the extension is over the square root of this length which thus goes as $N^{1/4} \ell_s$. This is what one expects from the free string picture at low string couplings. Note that in \cite{FRT} the effective temperature associated with decay of the $HES$ state also captures this $\sqrt{N}$ scaling of the size. At the correspondence point \cite{Horowitz:1996nw}, however the entire string should lie within its Schwarzchild radius which is $\ell_s$. This can only be explained through a non-perturbative analysis in the string coupling as carried out in \cite{Horowitz:1997jc} using a thermal-scalar formalism. In our tree-level analysis these non-perturbative effects are ignored, therefore we expect to match onto random walk expectations. \\
Using the distribution $\rho(r)$ we compute numerically the mean-square spread of the string. For a normalized distribution this is defined as $\vev{r^2} = \int_0^\infty dr \, r^2 \rho(r)$. Since we do not have access to the exact normalized distribution we consider a variant, a restriced-mean-square spread which we define as: 
\begin{align}
\vev{r^2}_{\texttt{R}}  &= \frac{ \int_{r_{\text{min}} }^{\texttt{R}} \, r^2\rho(r) dr } { \int_{r_{\text{min}} }^{r_{\text{max}}} \rho(r) dr } 
\label{eqr2}
\end{align}
Note, that since our saddle point evaluation limits us to start from a large enough $r_{\text{min}}>q_{\texttt{saddle}}^{-1} > 0$, we need to normalize the expectation value. The value $R$, with $r_{\text{max}} > R  > r_{\text{min}}$ parametrizes the string away from the centre of mass point, till where we look at the spread. For various values of $R$ we find that the restricted-mean-square spread fits to square root of $N$ scaling, consistent with the random walk picture. 
\begin{figure} 
\centering
{\includegraphics*[width= 0.5 \linewidth]{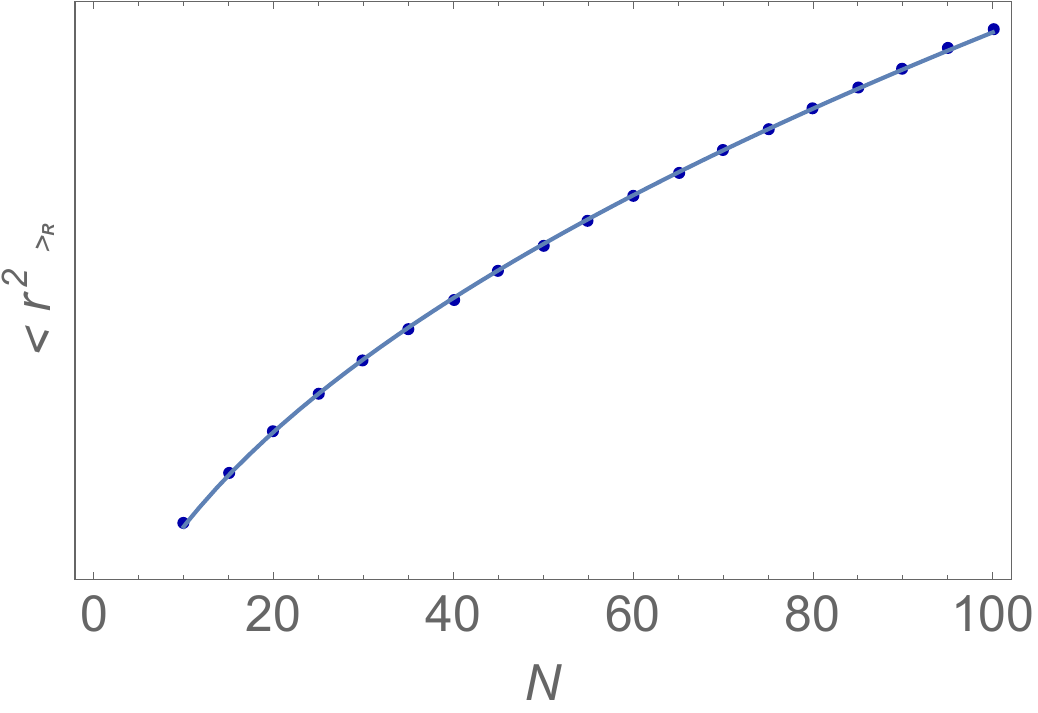}}
\caption{The above plot shows the second moment of $r$ has $\sqrt{N}$ scaling. The points are obtained from Eq.\eqref{eqr2}, and the line is the $\sqrt{N}$ fit. } \label{figr2}
\end{figure}

\section{Future directions}
\label{sec:future}
In this paper we have shown how tree-level bosonic string scattering amplitudes involving $\ket{HES}$ states contain features associated with quantum chaotic scattering, as well as quantum thermalization. Both of these features emerge statistically from the amplitudes. The statistical coarse-graining is facilitated via the large number of available microstates. The results lend support to the Susskind-Horowitz-Polchinski correspondence : the coarse-grained $\ket{HES}$ amplitudes are consistent with the thermal scales of the black hole at the correspondence point.  The non-coarse-grained amplitudes themselves are statistically chaotic in the random matrix theory sense. The chaotic nature plays a key-role, as it causes huge number of cancellations when the amplitudes are coarse-grained, resulting in consistency with the eigenstate thermalization hypothesis. We hope to make this point sharper in future investigations. Additionally, we will like to point out certain suggestive  directions:
\begin{itemize}
\item As indicated in the introduction, we will like to understand if the $HES$ S-matrix elements satisfy the non-trivial constraints as expected from a quantum gravity S-matrix which involves a black hole in the process \cite{Polchinski}. This stems from the operator statement \cite{Kiem:1995iy} : 
$$
a^\nd_{\tilde{h}'} a^\dagger_{h} = a^\dagger_{\tilde{h}} a^\nd_{h'},
$$
where $a_h$ are annihilation operators associated with near horizon modes. The tilde on the mode indicates a shift of the Kruskal time $u$: $\tilde{h}(u) = h(u-\Delta u)$ which occurs due to quantum backreaction at the horizon from the other mode. In the L.H.S the outgoing mode has this shift, while the shift in the ingoing mode of the RHS is not seen as this mode falls inside the horizon. The change in the Kruskal time is related to the delay in the Schwarzchild time coordinate $t$, which depends on the black hole parameters. It maybe possible to match to the shift at the correspondence point, through the different kinematical orderings in the $\vev{HHTT\cdots T}$ scattering amplitude. In this respect our result in Eq.\eqref{eq:HTdots-amp} will be useful.

\item Using the techniques of \cite{Eberhardt:2023xck} it should be possible to get the leading order $g_s$ corrections to the $\ket{HES}$ amplitudes, and understand what happens to the chaotic and the thermal features. Note, that at subleading orders in $g_s$, in the $AdS/CFT$ context, even the spectrum starts to show RMT statistics \cite{McLoughlin:2022jyt}. Going beyond the tree-level scattering will allow for: 
\begin{itemize}
\item Probing the Wigner time-delay \cite{de2002time}. This can be defined from the scattering amplitude in terms of energy derivative of the amplitude : $\Delta t = - {\cal A}^\dagger d {\cal A}/dE $. This time scale can be interpreted as the delay during the scattering process and is also called the {\em dwell time}. It will be interesting if this captures the Shapiro delay \cite{shapiro1964fourth} of the black hole at the correspondence point. In the analytic plane, the scale is set by the closest distance of the resonance pole off the real-axis, however at tree-level all the poles of ${\cal A}$ are destined to appear on the real energy axis, therefore it will be necessary to go to one loop. 
\item Another thing not visible in tree-level is how the long $\ket{HES}$ free string which follows the random walk picture gets squeezed into its own Schwarzchild radius, $r_s = \sqrt{\alpha'}$. As discussed in \cite{Horowitz:1997jc} this is a non-perturbative effect and using effective field theory they find: 
\begin{align}
r &= \frac{\sqrt{\alpha'}}{ g_s^2 \sqrt{N}}.
\end{align}
This relation is valid from $g_o < g_s < g_c$ where $g_o = N^{-3/8}$ and $g_c = N^{-1/4}$. Hence the string length $r$ interpolates from the random walk limit (at weak coupling) to the string scale at intermediate coupling. One will like to understand this calculation in the $\ket{HES}$ context from the $g_s$ corrections to the form factor computation. Recently the correspondence point was also discussed from this  point of view in the supersymmetric case in \cite{Chen:2021dsw}.
\item In the above analysis non-perturbative techniques of \cite{gross1987high} maybe useful. Another place where this maybe necessary will be to probe the chaotic fractal nature of the non-coarse-grained amplitude that is expected when $\ket{HES}$ states are involved. This is one of the conclusion in the absence of features from tree-level computations \cite{Hashi}.
\end{itemize}
\item The computation should be generalized to the case when the tachyon is replaced with photon. Due to extra polarization degrees of freedom available, by controlling different choices of $\zeta_i \cdot \zeta_j$ we might be able to speed-up / slow-down thermalization. Our work uncovered certain differences in the scales of chaos / thermalization when the $\ket{HES}$ polarizations were orthogonal / parallel to each other. It will be interesting to understand it completely including the role of the polarizations of the probe. 
\item The numerical precisions maybe pushed further in future to larger values of $N$ with smaller resolutions of scattering angles $\chi$ than considered in this work. In the context of off-diagonal ETH, Eq.\eqref{eq:eth}, the RMT behaviour ({\em uncorrelated} random numbers) is expected to hold only for small $\omega$. It turns out that there is a critical $\omega = \Delta E_{RMT}$ which is parametrically (in system size) smaller than the corresponding time scale for thermalization \cite{Richter:2020bkf, Wang:2021mtp}. Going to larger levels of the $\ket{HES}$ states by pushing the numerics, will allow us to explore this scalings of $\Delta E_{RMT}$ with $N$, which in the string context is a placeholder for system size.   
\item It is well known that due to the KLT relations \cite{kawai1986relation} closed string amplitudes can be expressed in terms of sum over open string amplitudes. Therefore using the tree-level results, one will be able to understand thermal and chaotic features of graviton scatterings within bosonic string theory. There is indication that near black hole horizons closed strings get stretched to open ones \cite{ab1,ab2,ab3} : the $\ket{HES}$ set-up can offer a way to realize this. There are also signatures of non-adiabatic dynamics involved in horizon crossing \cite{Silverstein:2014yza}, which will be interesting to realize in the $\ket{HES}$ context. 
\item Through the Mellin representation flat space scattering amplitudes are related to CFT correlators in Mellin space \cite{Penedones:2010ue}. Recently similar ideas have been used to obtain via bootstrapping methods, the $AdS$ string scattering amplitudes \cite{Gopakumar:2022kof}, which hints towards reorganizing an arbitrary $CFT_d$ amplitude as string scattering amplitude in $AdS_{d+1}$. A natural question is : {\em How the uplifted $\ket{HES}$ scattering amplitudes in $AdS$ are codified into a $CFT$ correlation function?} It will be interesting to see if the answer to this question is related to the recently discussed OPE randomness hypothesis \cite{Belin1, Belin2, Belin3, Belin4} which defines random CFTs. Features of higher dimensional gravity \eg $\,$ wormholes are known to emerge from these random CFTs \cite{Maloney:2020nni, Cotler:2020ugk, Chandra1}. 
\item S-matrix elements have been computed for black holes realized in the intersection of D1-D5 systems \cite{Lunin:2012gz}. It will be interesting to understand how chaos appears in this set-up. 
\item Recently \cite{Ceplak:2023afb} pointed out certain puzzles and their resolutions modulo the extremal Kerr case in the context of the black hole / string correspondence. One could explore what the $\ket{HES}$ string amplitudes can contribute towards these discussions.
\item  Recently \cite{Pesando}  attributes the chaotic origins of the amplitudes to the chaoticity of the coefficients present in the decomposition of a large level DDF state into the $SO(D-1)$ irreps. It will be interesting if this can explain the appearence of GOE statistics for $HHT$ and GUE for $HTT$. 
\end{itemize}

%
%
%

\section*{Acknowledgement}
It is a pleasure to thank Arjun Bagchi, Sumit R. Das, Shouvik Datta, Justin David, Lorenz Eberhardt, Apratim Kaviraj, Nilay Kundu, Juan Maldacena, Sridip Pal and Arnab Sen for useful discussions. DD will like to thank the participants of the December 2023 pre-ISM workshop at TIFR, India; the audience at University of Crete, University of Kentucky and University of California Los Angeles, where this work was presented. AS would like to thank the participants of ISM 2023 at IIT-Bombay, where a poster on this work was presented. The authors would like to acknowledge the support provided by the Max Planck Partner Group grant MAXPLA/PHY/2018577 and {{SERB/PHY/2020334}}. 

%
%
%
%
%
%
%
%
\newpage 		
	
\appendix

\renewcommand{\theequation}{\Alph{section}.\arabic{equation}}
 
 \begin{centering}
 \section*{Appendices}
 \end{centering}

 \section{Kinematics } 
 \label{app:kin}
 Here we explicitly specify our kinematic choice for both the three point as well as four point amplitude computations. 
 \subsection{Three point kinematics}
 	\subsubsection*{String States Final Momenta}
	The list of constraints are, 
	\begin{align}
			&\left (p_1 \right )^2 = -M_1^2 = -2( N_1-1), \quad 
			\left (p_2 \right )^2 = -M_2^2 = -2( N_2-1),\nn \\
			&\left (p_3 \right )^2 = 2, \quad
			p_1+p_2+p_3 = 0.
		\label{constraint}%
	\end{align}
	Maintaining the constraints we choose:
	\begin{align}
			p_1 &= \sqrt{2 (N_1-1)} (1,0,0),\,\,\,
			p_2 = - \sqrt{2 (N_2-1)} \left ( 
			\frac{N_1+N_2-1}{\sqrt{2(N_1-1)}\sqrt{2(N_2-1)}},
			\kappa \cos \theta, \kappa \sin \theta \right )\quad \nn\\
			p_3 &= -\left ( 
			\frac{N_1-N_2-1}{\sqrt{2(N_1-1)}},
			- \kappa \sqrt{2(N_2-1)} \cos \theta,
			- \kappa \sqrt{2(N_2-1)} \sin \theta
			\right ), \label{momenta} \\ 
			&\text{where} \;\kappa = 
			\sqrt{\frac{(N_1+N_2-1)^2}{2(N_1-1)2(N_2-1)}-1}. \nn
	\end{align}

	\subsubsection*{DDF Photon Momenta}
We take the photons of both the $\ket{HES}$ to be 
	parallel i.e. $q_1\propto 
	q_2$, i.e, $q_1 \equiv q, ~
	q_2 \equiv -\gamma q$. Temporal component of DDF photon associated with $\ket{HES_2}$ is negative, which is fixed by the sign of $\gamma$. Furthermore we also need $q_1^2 = 0$ and $q\cdot \tilde{p}_1 = 1$. These are satisfied by: 
	\begin{equation} \label{eqn q}
		\begin{split}
			& q = -\frac{1}{\sqrt{2 (N_1-1)}} \left(1,\cos (\delta),\sin (\delta )\right) \\
			\text{where, } & \delta = \theta + \sec ^{-1}\left(\frac{-\gamma  \sqrt{(N_1-N_2)^2+2 (N_1+N_2)-3}}{(2-\gamma ) N_1-\gamma  (N_2-1)-2}\right)
		\end{split}
	\end{equation}
We prove in the following that $\gamma$ is strictly positive. Writing out explicitly the condition: $q_2 \cdot \tilde{p}_2 = - \gamma q_1 \cdot \tilde{p}_2= 1$,  we obtain:	
	\begin{align}
			&\gamma \left( \frac{\sqrt{N_1^2-
					2 N_1 (N_2-1)+N_2^2+2 N_2-3}}{2 (N_1-1)}  
			\cos (\delta -\theta )+\frac{N_1+N_2-1}
			{2 (N_1-1)} \right) =1. \label{eq:gamma}
			\end{align}
When, $N_1,N_2 \gg 1$ then using $N_1^2-2 N_1 (N_2-1)+N_2^2+2 N_2-3 = (N_1-N_2)^2 + 2(N_1+N_2) -3$ we get : 
			\begin{align} 
			(N_1-N_2)^2 + 2(N_1+N_2) -3 
			&\ll (N_1+N_2-1)^2, \nn \\
			\text{or, }\,\, (N_1+N_2-1) \gg \sqrt{(N_1-N_2)^2 + 2(N_1+N_2) -3} 
			\;\ \cos (\delta - \theta) &\gg -(N_1+N_2-1). 
			\end{align}
Therefore for the first equation in Eq.\eqref{eq:gamma} this fixes: 
			\begin{align}
			 \left(\frac{\sqrt{N_1^2-
					2 N_1 (N_2-1)+N_2^2+2 N_2-3}}{2 (N_1-1)}\right)  
			\cos (\delta -\theta )+\frac{N_1+N_2-1}
			{2 (N_1-1)} & > 0.
		\end{align}
From Eq.\eqref{eq:gamma} we find that $\gamma$ has the same sign as the l.h.s above, and therefore is positive. Also note that in the limit of large $N_1= N_2= N$, the variable $\gamma$ takes the form 
	\begin{equation}
		\begin{split}
			\gamma&= 1+ \sqrt{\frac{1}{{N}}} \cos (\delta-\theta )+\frac{\cos (2 \delta - 2\theta )}{{N}}+O\left(\left(\frac{1}{{2N}}\right)^{3/2}\right) \simeq 1
		\end{split}
	\end{equation}

 Next, recall for DDF states : $\zeta = \lambda - ( p \cdot \lambda ) q$. In what follows we work with two choices of polarization inner products. 
\begin{itemize}
\item { $\mathbf{ \lambda_1 \cdot \lambda_2 = 0}$} : 
For the photon polarization, we take $\lambda_2 = -\lambda_1 = -\lambda$, which satisfies $ q.\lambda = 0, ~ \lambda^2 = 0$. Then $\lambda$ can be written in terms of $\delta$ as : 
	$$ \lambda = \frac{1}{\sqrt{2}} \left( 0, \sin(\delta), -\cos(\delta), i \right).$$
	Note that for this choice of polarization $\zeta_1 \cdot \zeta_2 =0$ and thus this choice hides the effect of the derivative contractions among the $\ket{HES}$ states. 

\item { $\mathbf{\lambda_1 \cdot \lambda_2 \neq 0}$} :  The other choice of polarizations maintaining $\lambda_1^2 = \lambda_2^2 = \lambda_1 \cdot q_1 = \lambda_2 \cdot q_2 = 0$ is : 
	\begin{align}
		\lambda_1 &=  \frac{1}{\sqrt{2}} \left( 0, \sin(\delta), -\cos(\delta), i \right), \quad
		\lambda_2 = \frac{1}{\sqrt{2}} \left( 0, \sin(\delta), -\cos(\delta), -i \right).
	\end{align} 
	Note that automatically $\lambda_1 \cdot q_2 = \lambda_2 \cdot q_1 = 0$ is also guaranteed. And now, the dot product gives: 
	\begin{align}
		\zeta_1 \cdot \zeta_2 &= \lambda_1\cdot \lambda_2 = 1.
	\end{align}
\end{itemize}

\subsection{Four point kinematics}
\label{app:kin2}

For the $HHTT$ amplitude, the list of constraints are, 
	\begin{align}
			\left (p_1 \right )^2 &= -M_1^2 = -2( N_1-1), \,\,
			\left (p_2 \right )^2 = 2, \,\,
			\left (p_3 \right )^2 = -M_2^2 = -2( N_3-1), \nn \\ 
			\left (p_4 \right )^2 &= 2, \,\,
			\vec{p_1}+\vec{p_2}=0=\vec{p_3}+\vec{p_4}, \,\,\,
			p_1+p_2+p_3+p_4 = 0.
		\label{constraints}%
	\end{align}
	Maintaining the constraints we get:
	\begin{align}
			&p_1=\left\{\sqrt{e_0^2+2 N_1},\sqrt{e_0^2+2} 
			\sin (\theta_1),\sqrt{e_0^2+2} \cos (\theta_1),0\right\}, \,\,
			p_2=\left\{e_0,-\sqrt{e_0^2+2} \sin (\theta_1),-\sqrt{e_0^2+2} \cos (\theta_1),0\right\}, \nn \\ 
			&p_3=\left\{-\sqrt{e_0'^2+2 N_3},\sqrt{e_0'^2+2} \sin (\theta_2 ),\sqrt{e_0'^2+2} \cos (\theta_2 ),0\right\}, \nn \\
			&p_4=\left\{-e_0',-\sqrt{e_0'^2+2} \sin (\theta_2 ),-\sqrt{e_0'^2+2} \cos (\theta_2 ),0\right\},	\label{momenta}
		\end{align}
where $e_0'= \frac{-N_3 \sqrt{e_0^2+2 N_1}+N_1 \sqrt{e_0^2+2 N_1}+e_0 N_1+e_0 N_3}{2 N_1}$. 

\ndt The DDF photon momenta are given by : 	
	\begin{align}
			q_1&=\left\{\frac{-1}{\sqrt{e_0^2+2 N_1}+e_0},\frac{\left(e_0^2+2\right)^{-1/2}e_0 \sin (\theta_1)}{ \left(\sqrt{e_0^2+2 N_1}+e_0\right)},\frac{e_0 \cos (\theta_1)}{\sqrt{e_0^2+2} \left(\sqrt{e_0^2+2 N_1}+e_0\right)},\frac{\sqrt{2}}{\sqrt{e_0^2+2} \left(\sqrt{e_0^2+2 N_1}+e_0\right)}\right\}\nn \\
			q_3&= -\gamma q_1 \;\ \text{where} \;\ \gamma= - \frac{1}{p_3 \cdot q_1} = \frac{\sqrt{e_0^2+2} \left(\sqrt{e_0^2+2 N_1}+e_0\right)}{\sqrt{e_0^2+2} \sqrt{e_0'^2+2 N_3}-\sqrt{e_0'^2+2} e_0 \cos (\theta_1-\theta _2)}.
		\end{align}
	For this particular choice of kinematics, the polarization vectors of the DDF photons must contain 
	an imaginary $5$-th component (to make the polarization vectors to be null-like) :
	
	\begin{align}
			\lambda_1&=\frac{1}{\sqrt{2}}\{0,\cos \theta_1,-\sin \theta_1,0,i\},  \lambda_3=\frac{1}{\sqrt{2}}\{0,\cos \theta_1,-\sin \theta_1,0,-i\}, 
			\zeta_1=\frac{1}{\sqrt{2}}\{0,\cos \theta_1,-\sin \theta_1,0,i\} \nn \\ 
			\zeta_3&=\frac{1}{\sqrt{2}}\bigg\{\frac{\sqrt{(e_0'^2+2)(e_0^2+2)} \sin (\theta_1-\theta_2)}{\sqrt{e_0^2+2} \sqrt{e_0'^2+2 N_3}-\sqrt{e_0'^2+2} e_0 \cos (\theta_1-\theta_2)},\frac{\sqrt{(e_0^2+2)(e_0'^2+2 N_3)}\cos (\theta_1)-\sqrt{e_0'^2+2} e_0 \cos (\theta_2)}{\sqrt{e_0^2+2} \sqrt{e_0'^2+2 N_3}-\sqrt{e_0'^2+2} e_0 \cos (\theta_1-\theta_2)},\nn \\
			& \;\ \;\ \;\ \;\ \;\ \;\ \;\ \;\ \frac{-\sqrt{(e_0^2+2)(e_0'^2+2 N_3)}\sin (\theta_1)+\sqrt{e_0'^2+2} e_0 \sin (\theta_2)}{\sqrt{e_0^2+2} \sqrt{e_0'^2+2 N_3}-\sqrt{e_0'^2+2} e_0 \cos (\theta_1-\theta_2)},0,-i\bigg\}.
		\end{align}
If we consider the case $N_1=N_3=N$, which corresponds to $e_0=e_0'$,  the kinematics and hence the amplitude itself becomes simple:
	\begin{align}
		p_1&=\left\{\sqrt{e_0^2+2 N},\sqrt{e_0^2+2} 
			\sin (\theta_1),\sqrt{e_0^2+2} \cos \theta_1,0\right\}, \,\,
			p_2=\left\{e_0,-\sqrt{e_0^2+2} \sin \theta_1,-\sqrt{e_0^2+2} \cos \theta_1,0\right\} \nn \\
			p_3&=\left\{-\sqrt{e_0^2+2 N},\sqrt{e_0^2+2} \sin \theta_2 ,\sqrt{e_0^2+2} \cos \theta_2 ,0\right\}, \,
			p_4=\left\{-e_0,-\sqrt{e_0^2+2} \sin \theta_2 ,-\sqrt{e_0^2+2} \cos \theta_2 ,0\right\}\nn \\
			q_1&=\left\{-\frac{1}{\sqrt{e_0^2+2N}+e_0},\frac{e_0\left( e_0^2 + 2 \right)^{-1/2} \sin \theta_1}{ \left(\sqrt{e_0^2+2 N}+e_0\right)},\frac{e_0 \cos \theta_1}{\sqrt{e_0^2+2} \left(\sqrt{e_0^2+2 N}+e_0\right)},\frac{\sqrt{2}}{\sqrt{e_0^2+2} \left(\sqrt{e_0^2+2N}+e_0\right)}\right\}\label{eq:kin4}\\
			q_3&= -\gamma q_1 \;\ ; \gamma= \frac{\sqrt{e_0^2+2 N}+e_0}{\sqrt{e_0^2+2 N}-e_0 \cos (\theta_1-\theta_2)}.\nn
		\end{align}

\section{Details of $P(a_i,b_i)$ and $Q_j(l_i)$}
\label{app:PQ}
Here we give the explicit forms of the factors present in the evaluated amplitudes, \eg Eq.\eqref{HHT amp} and Eq.\eqref{eq:amp4amp3}: 
	\begin{align}
		P(a_i,b_i)&= \sum_{m_1,m_2=1}^{a_i,b_i}  (-1)^{m_2 +1 }\bigg[   (\zeta_1 \cdot \zeta_2)\frac{(m_2 + m_1 -1)!  }{(m_1 - 1)! (m_2 - 1)!  } S_{a_i-m_1} \bigg( -\frac{a_i}{s_1\gamma} \bigg) S_{b_i-m_2} \bigg( -\frac{b_i (-1)^{s_2}}{s_2} \gamma \bigg)  \bigg] \label{eq:p} \\
		Q_1(l_1)&= \sum_{m_1=1}^{l_1}   (\zeta_1 \cdot p_2) S_{l_1-m_1} \bigg( -\frac{l_1}{s_1\gamma} \bigg)=(\zeta_1 \cdot p_2)V_1(l_1) \label{q1}\\
		Q_2(l_2)&= \sum_{m_2=1}^{l_2}   (\zeta_2 \cdot p_1)(-1)^{m_2} S_{l_2-m_2} \bigg( -\frac{l_2 (-1)^{s_2}}{s_2} \gamma \bigg)= (\zeta_2 \cdot p_1)V_2(l_2) \label{q2}
	\end{align}
	Next, we note that the Schur polynomials can now be expressed in terms of products :
	\begin{align}
		S_{r_1 - m_1} &= \frac{ (-1)^{r_1 - m_1 } }{(r_1-m_1)!} \prod_{s_1=1}^{r_1-m_1} \bigg( \frac{r_1}{\gamma} - s_1 +1 \bigg), \,\, 
		S_{r_2 - m_2} = \frac{1}{(r_2-m_2)!} \prod_{s_2=1}^{r_2-m_2} \bigg( r_2 \gamma - s_2 +1 \bigg) .
	\end{align}
Hence, we can simplify Eq.\eqref{eq:p} as: 
	\begin{align}
	P(a_i,b_i)  &=    \sum_{m_1,m_2=1}^{a_i,b_i}\frac{ (-1)^{m_2 +1+a_i - m_1 }}{(a_i-m_1)! (b_i- m_2)!}\bigg[   \frac{(m_2 + m_1 -1)!  }{(m_1 - 1)! (m_2 - 1)!  } \bigg] \prod_{s_1=1}^{a_i-m_1} \bigg( \frac{a_i}{\gamma} - s_1 +1 \bigg) \nonumber \\
		&\times  \prod_{s_2=1}^{b_i-m_2}  \bigg( b_i \gamma - s_2 +1 \bigg) \nn \\
		&=(-1)^{a_i} \prod_{j_1=1}^{a_i-1} \left( \frac{a_i/\gamma- a_i  + j_1}{j_1}\right) \prod_{j_2=1}^{b_i-1}\bigg( \frac{b_i \gamma - b_i +j_2}{j_2}\bigg) \bigg( \frac{(\gamma-1)a_i b_i}{a_i -b_i \gamma } \bigg) \nn \\
		&= (-1)^{a_i} \prod_{j_1=1}^{a_i-1} \left( \frac{a_i/\gamma-  j_1}{j_1}\right) \prod_{j_2=1}^{b_i-1}\bigg( \frac{b_i \gamma - j_2}{j_2}\bigg) \bigg( \frac{(\gamma-1)a_i b_i}{a_i - b_i \gamma } \bigg) .
	\end{align}
Similarly, carrying out the sums in equation \eqref{q1} and equation \eqref{q2} yields simpler product formula expressions of $Q_1$ and $Q_2$,
		\begin{align}
			Q_1(l_1)&= (\zeta_1 \cdot p_2)(-1)^{l_1} \prod_{j_1=1}^{l_1} \left(\frac{l_1/ \gamma-j_1}{j_1}\right) \nonumber \\
			&=-\left(\sqrt{\frac{N_1+N_2-1}{ \gamma}-\frac{N_1-1}{\gamma^2}-(N_2-1)}\right)(-1)^{l_1} \prod_{j_1=1}^{l_1} \left(\frac{l_1/ \gamma-j_1}{j_1}\right)\\
			Q_2(l_2)&= (\zeta_2 \cdot p_1)\prod_{j_2=1}^{l_2} \left(\frac{l_2 \gamma-j_2}{j_2}\right)\nn \\
			&=-\left(\sqrt{(N_1+N_2-1) \gamma-(N_1-1)-(N_2-1)\gamma^2}\right) \prod_{j_2=1}^{l_2} \left(\frac{l_2 \gamma-j_2}{j_2}\right).
		\end{align}

\section{Unfolding procedure}
\label{app:unfolding}

\begin{figure}[h]
	\centering
	\includegraphics[width=.7\linewidth]{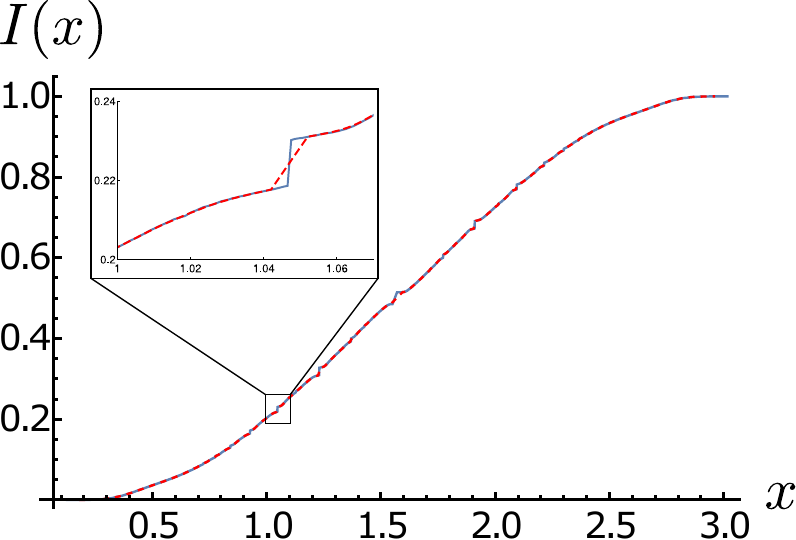}  
	\caption{ $I(x)$ for the peak distribution in three-point amplitude, shown in blue. $I_{av} (x)$, shown as the red dashed curve, is obtained by performing a running spectral average over 10 consecutive peaks. }
	\label{fig:unfolding}
\end{figure}

To study the random matrix feature of the level spacing distribution of a spectrum, one needs to separate the local fluctuation of the levels from the overall energy dependence of the level separations. To do this, we follow the procedure discussed in \cite{ABULMAGD2014185}. From a given ensemble of peak positions or levels $ \{ x_1, x_2, \cdots \, x_M \} $, we first compute the cumulative spectral function $I(x)$, which is defined as:
\begin{align}
	I (x) = \frac{ \text{number of peak positions} \leq x}{\text{total number of peaks}} = \frac{\# \text{ of }x_i \leq x }{M}.
\end{align}
This is also known as the staircase function. This can be then decomposed as: $I (x) = I_{av} (x) + I_{fluc}(x)$, where the derivative of $I_{av} (x)$ produces the mean level density $\rho (x)$ and $I_{fluc}(x)$ denotes the fluctuations about this mean distribution. Thus, for a known $\rho(x)$, 
\begin{align}
	I_{av} (x) = \int_{- \infty}^{x} dy \, \rho (y).
\end{align}
$I_{av} (x)$ can also be computed by performing a running spectral average on the level distribution. In our case, as the mean level distribution is not known apriori, we compute $I_{av} (x)$ using this running average method. See Fig. \ref{fig:unfolding} for unfolding the peaks in a sample case with $N=30$ DDF state. Once the $I_{av} (x)$ is obtained, one can now define the unfolded spectrum by introducing dimensionless energy variable:
\begin{align}
	u_i = I_{av} (x_i).
\end{align}
The unfolded level spacing distribution is obtained from the consecutive level spacings of these $u_i$ variables.

\section{Total number of $HES$ states}
\label{app:omega}
In principle one HES might be constructed from any arbitrary number of DDF photons with arbitrary momenta. But for the simplicity of the calculation we are restricted to the case where all the constituent DDF photon momenta are parallel ($\sim n q$ with different integer mode number $n$). The polarization of these photons can also be different. In a $D$-dimensional spacetime, each polarization vector can be linear combination of $D-2$ linearly independent null like vectors. For the process involving $2$ HES states, if we put a constraint $\lambda_1 \cdot \lambda_2=$constant, then the number of choice of independent polarizations for $\lambda_1$ is $D-2$, and for polarization vector $\lambda_2$ is $D-3$, due to the constraint. the total number of microstates is given by,
	
	\begin{align}		
			\Omega &= \Omega(N_1) \Omega(N_2)
			=\underset{\lbrace n_{r_1}\rbrace}{\sum} \underset{\lbrace n_{r_2}\rbrace}
			{\sum}\delta(N_1-\sum_{r_1}r_1 n_{r_1})\delta(N_2-\sum_{r_2}r_2 n_{r_2})
			\prod_{r_1} C_{r_1} \prod_{r_2} C_{r_2},
	\end{align}
	where : 
	$$
	C_{r_1} = \left(\begin{matrix}
				D-2+n_{r_1}-1\\
				n_{r_1}
			\end{matrix} \right) = (-1)^{n_{r_1}} \left(\begin{matrix}
				2-D\\
				n_{r_1}
			\end{matrix} \right), \,\,
			C_{r_2} = \left(\begin{matrix}
				D-3+n_{r_2}-1\\
				n_{r_2}
			\end{matrix} \right) = (-1)^{n_{r_2}} \left(\begin{matrix}
				3-D\\
				n_{r_2}
			\end{matrix} \right).
	$$
Therefore we have:
\begin{align}
			\Omega(N_1) \Omega(N_2) &= 
			\int d \beta_1 d \beta_2 ~ e^{\beta_1 N_1 + \beta_2 N_2} 
			\prod_{r_1} \left(e^{-\beta_1 r_1} \right)^{n_{r_1}}
			\left(\begin{matrix}
				2-D\\
				n_{r_1}
			\end{matrix} \right) 
			\prod_{r_2} \left(e^{-\beta_2 r_2} \right)^{n_{r_2}} 
			\left(\begin{matrix}
				3-D\\
				n_{r_2}
			\end{matrix} \right)\nn \\
		&\overset{N_1, N_3 \rightarrow \infty}{\sim} \exp \bigg[ \frac{2\pi}{\sqrt{6} } \left( \sqrt{ N_1(D-2) } + \sqrt{N_3 ( D-3 ) } \right)\bigg]. \label{omega12}
	\end{align}

\subsection{Number of states with fixed polarizations}

     In our calculation we have chosen the DDF photon momenta of both the $\ket{HES}$ states to be parallel. Now, we also fix the polarization vectors of these photons:  $\ket{HES_1} (\ket{HES_2})$ photons having polarization $\lambda_1 (\lambda_2)$ which get fixed kinematically. The number of microstates in the process involving two $\ket{HES}$ states (levels $N_1$ and $N_2$ respectively) is then given by, 
	\begin{align}
			\Omega &= \Omega(N_1) \Omega(N_2)
			=\underset{\lbrace n_{r_1}\rbrace}{\sum} \underset{\lbrace n_{r_2}\rbrace}
			{\sum}\delta(N_1-\sum_{r_1}r_1 n_{r_1})\delta(N_2-\sum_{r_2}r_2 n_{r_2}).
	\end{align}
	Next using Lagrange multipliers $\beta_1 $ and $\beta_2$ at large $N_1, N_2$ we obtain: 		
	\begin{align}
			\Omega(N_1) \Omega(N_2) &\sim \exp \left[ \frac{2 \pi}{\sqrt{6}} \left( \sqrt{N_1}+\sqrt{N_2} \right)\ \right].
	\end{align}

\section{Generalization to, $HES_1 + T_1 +\cdots  + T_{n-k} \rightarrow HES_2 + T_{n-k+1} + \cdots + T_n  $ amplitude}
\label{app:higher} 
Let's consider the process involving $2$ HES states $H_1$ and $H_2$ and many ($n-2$) tachyons ($T_3,T_4 \ldots T_n$). The notion of ingoing and outgoing is not important here, that is encoded in the explicit form of the kinematic momenta. The amplitude can be written as, 

\begin{equation}
	\begin{split}
		\mathcal{A}&=\frac{1}{Vol(SL_2)}\int \prod_{i=1}^{n} e^{\cal L}\\
	\end{split}
\end{equation}
\begin{align}
	e^{\cal L} =\bigg\langle :\prod_{r_1} &\left( \sum_{m_1}^{r_1} \frac{i }{(m_1 - 1)!} \zeta_1 \cdot \partial^{m_1} X_1 S^1_{r_1-m_1} \right) e^{i p_1 \cdot X_1}:: \prod_{r_2} \left( \sum_{m_2}^{r_2} \frac{i }{(m_2 - 1)!} \zeta_2 \cdot \partial^{m_2} X_2 S^3_{r_2-m_2} \right) \nonumber \\ &e^{i p_2 \cdot X_2}: : e^{i p_3 \cdot X_3 }: : e^{i p_4 \cdot X_4 }: \ldots : e^{i p_n \cdot X_n}:\bigg\rangle
\end{align}
For the sake of simplicity of the calculation we make kinematic choices : 
\begin{align}
	q_2 &= -\gamma q_1, \,\,\, \zeta_1 \cdot \zeta_2 = \lambda_1 \cdot \lambda_2, \\
	\zeta_i \cdot q_j &= 0, \,\,\, \lambda_i^2 = 0.
\end{align}
Carrying out the possible contractions yield:  

\begin{equation}
	\begin{split}
		\sum_{k=0}^{\lbrace J_1,J_2 \rbrace_{min}} \frac{1}{k!} \sum_{\substack{\lbrace a(k)\rbrace \subset \lbrace r_{1} \rbrace\\ \lbrace b{(k)}\rbrace \subset \lbrace r_{2} \rbrace\\ \text{ordered}}} &\prod_{i=1}^{k} \Bigg[ \sum_{m_1,m_2=1}^{a_i,b_i} \bigg \lbrace \frac{(-1)^{m_2+1}(m_1+m_2-1)! \zeta_1.\zeta_2}{(m_1-1)!(m_2-1)! z_{21}^{m_1+m_2}}  \bigg \rbrace \\
		&\times S_{a_i-m_1} \bigg( \frac{a_i}{s_1} \left( \sum_{i \neq 1} \frac{q_1 \cdot p_i}{z_{i1}^{s_1} } \right) \bigg) S_{b_i-m_2} \bigg( \frac{b_i}{s_2} \bigg(  \sum_{i \neq 2}\frac{q_2 \cdot p_i}{z_{i2}^{s_2} }  \bigg) \bigg) \Bigg]\\
		\times &\prod_{l_1 \in \overline{\lbrace a(k) \rbrace}} \Bigg[ \sum_{n_1=1}^{l_1}\bigg( \frac{ \zeta_1 \cdot p_2 }{z_{21}^{n_1} }+\frac{ \zeta_1 \cdot p_3 }{z_{31}^{n_1} } \ldots  +\frac{ \zeta_1 \cdot p_n }{z_{n1}^{n_1} }  \bigg)   S_{l_1-n_1} \Bigg]\\ 
		\times 
		&\prod_{l_2 \in \overline{\lbrace b(k) \rbrace}} \Bigg[ \sum_{n_2=1}^{l_2}\bigg(\frac{ \zeta_2\cdot p_1 }{z_{12}^{n_2} } +  \frac{ \zeta_2 \cdot p_3 }{z_{32}^{n_2} } \ldots +  \frac{ \zeta_2 \cdot p_n}{z_{n2}^{n_2} }  \bigg) S_{l_2-n_2} \Bigg] \times \prod_{i<j} z_{ij}^{p_i \cdot p_j }.
	\end{split}
	\label{eq:HTdots-amp}%
\end{equation}

\section{Norm of the $\ket{HES}$ states} 
\label{app:norm}

To calculate the norm of the DDF states, following \cite{GR} we use the commutation relation of DDF operators which is same as the creation operators:
\begin{align}
	\left[\alpha^\mu_m, \alpha^\nu_n \right] = m \delta_{m,-n} \eta^{\mu, \nu}  \Rightarrow  \left[ \lambda_1 \cdot A_m, \lambda_2 \cdot A_n \right] = m \lambda_1 \cdot \lambda_2 \delta_{m,-n} .
\end{align}
Thus for a state with single creation operator the norm will be:
\begin{align}
	| | \lambda \cdot A_{-m_1} |0 \rangle |^2  =  \vev{0 | \lambda^* \cdot A_{m_1} \lambda \cdot A_{-m_1} | 0 } = m_1 |\lambda|^2 = m_1 ,
\end{align}
where $|\lambda| = 1$. Thus for a generic DDF state we can write:
\begin{align}
	| |\lambda \cdot A_{-m_1} \cdots \lambda \cdot A_{-m_n} | 0 \rangle |^2 &= \vev{0 | \lambda^* \cdot A_{m_1} \cdots \lambda^* \cdot A_{m_n} \lambda \cdot A_{-m_n} \cdots \lambda \cdot A_{-m_1} | 0 }  \nn \\
	&\equiv  \vev{0 | A_{m_1} \cdots  A_{m_n}  A_{-m_n} \cdots  A_{-m_1} | 0 }.
\end{align}
For a state with two creation operators, we can use commutations to obtain:
\begin{align}
	| |\lambda \cdot A_{-m_2}  \lambda \cdot A_{-m_1} | 0 \rangle |^2 =  \vev{0 | A_{m_1}  A_{m_2}  A_{-m_2}  A_{-m_1} | 0 } &= m_1 m_2 \left( \delta_{m_1,m_2} \delta_{m_2,m_1} + \delta_{m_1,m_1} \delta_{m_2,m_2} \right)  \nn \\
	&=  m_1^2 \delta_{m_1,m_2} +  m_1 m_2  .
\end{align}
To find a general expression for DDF norms, we first look into the case for three creation operators:
\begin{figure}[h]
	\centering
	\includegraphics[scale=0.9]{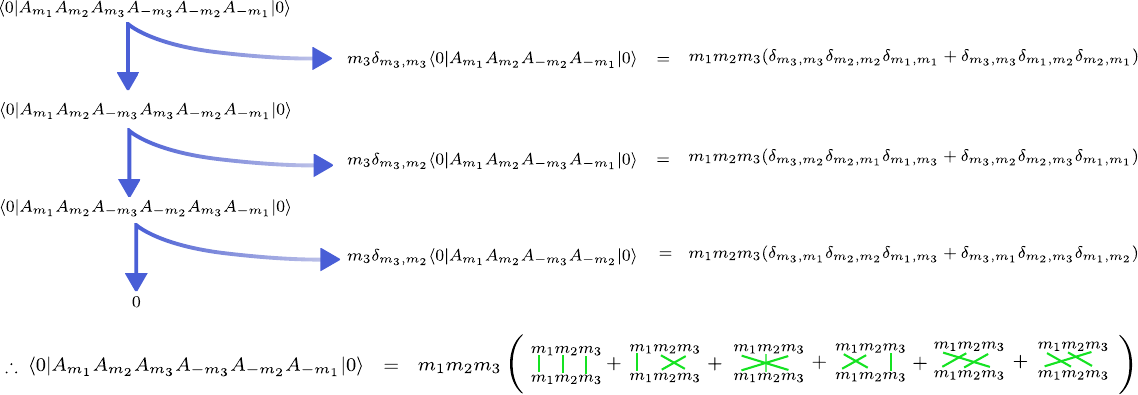}
	\label{figDDFNorm}
\end{figure}
\ndt In the expression below, the green lines indicate the pair of indices appearing in delta functions, and the two sets of indices appear for the two sets of creation and annihilation operators. This amounts to keeping one set of indices fixed and pair all the $3!$ permutations of the other set of indices and pairing them position-wise. Each term in the expression contains all 3 indices, and collectively the expression in parentheses count the number of same indexed pairs that one can make out of $\{m_1, m_2, m_3\}$.

\ndt For example, for $\{m_1, m_2, m_3\} = \{1, 2, 4\}$ the only set of pair one can create is $\{(1,1), (2,2), (4,4)\}$, so only the first term contributes and the norm squared is $m_1 m_2 m_3 \times 1$. But for $\{m_1, m_2, m_3\} = \{1, 3, 3\}$ the first and second terms in the parentheses contribute, thus the norm squared will be $m_1 m_2 m_3 \times 2 = m_1 \times m_2^2 \times 2!$. And for all three indices same one gets $m_1 m_2 m_3 \times 3! = m_1^3 \times 3!$.

\ndt Following this, we can generalize this procedure for any arbitrary number of creation operators: as one commutes an annihilation operator through the creation operators to bring it to the rightmost end, it produces a delta function for each commutation. Thus at the end one gets a sum of product of delta functions which amounts to counting the number of same indexed pairs from the annihilation and creation operators. So for a state created with $n_{m_i}$ number of $A_{-m_i}$ creation operators, the norm squared comes out to be:
\begin{align}
	\mathcal{N}^2 =  \bigg| | \underbrace{A_{-m_k} \cdots A_{-m_k}}_{n_{m_k}}  \cdots \underbrace{A_{-m_1} \cdots A_{-m_1}}_{n_{m_1}}  | 0 \rangle \bigg|^2  =  \prod_i^k m_i^{n_{m_i}} \cdot n_{m_i}!
\end{align}


\providecommand{\href}[2]{#2}\begingroup\raggedright\endgroup

\end{document}